\documentclass[a4paper,preprint,superscriptaddress,nofootinbib,longbibliography]{revtex4-1}
\usepackage[dvipdfmx]{graphicx}
\usepackage{bm}
\usepackage{amssymb}
\usepackage{amsmath}
\usepackage{color}
\usepackage{cancel}
\usepackage{comment}
\usepackage{here}
\usepackage{hyperref}

\def\beq#1\eeq{\begin{equation}#1\end{equation}}
\def\bal#1\eal{\begin{align}#1\end{align}}

\begin{document} 

\title{Electroweak baryogenesis in the three-loop neutrino mass model with dark matter}

\author{Mayumi Aoki}
\email{mayumi.aoki@staff.kanazawa-u.ac.jp}
\affiliation{Institute for Theoretical Physics, Kanazawa University, Kanazawa 920-1192, Japan}

\author{Kazuki Enomoto}
\email{k\_enomoto@kaist.ac.kr}
\affiliation{Department of Physics, KAIST, Daejeon 34141, Korea}

\author{Shinya Kanemura}
\email{kanemu@het.phys.sci.osaka-u.ac.jp}
\affiliation{Department of Physics, Osaka University, Toyonaka, Osaka 560-0043, Japan}


\preprint{KANAZAWA-22-08, OU-HET-1165}

\begin{abstract}
\vspace*{1pt}
Baryon asymmetry of the Universe is evaluated in the model originally proposed in Phys. Rev. Lett. 102 (2009) 051805, where Majorana masses of neutrinos are generated via three-loop diagrams composed of additional scalar bosons including the dark matter candidate which is odd under an unbroken $Z_2$ symmetry. In order for the model to include multiple CP-violating phases, we do not impose the softly broken $Z_2$ symmetry imposed in the original model to avoid the flavor-changing neutral current at tree level. Instead, for simplicity, we assume the flavor alignment structure in the Yukawa interactions. We also simply assume the alignment structure in the Higgs potential so that the Higgs couplings coincide with those in the SM at tree level. Under these phenomenological simplifications, the model still contains multiple CP-violating phases. By using destructive interferences among them, it is compatible with the stringent constraint from the electric dipole moment measurements to generate the observed baryon asymmetry along with the scenario of electroweak baryogenesis. We show a benchmark scenario which can explain neutrino mass, dark matter and baryon asymmetry of the Universe simultaneously and can satisfy all the other available experimental data. Some phenomenological predictions of the model are also discussed.
\end{abstract}

\maketitle


\section{Introduction}

It is one of the most important tasks in high energy physics to explain phenomena that cannot be explained in the standard model (SM), such as neutrino oscillation~\cite{Super-Kamiokande:1998kpq, ref: SNO}, dark matter (DM)~\cite{Planck:2018vyg} and baryon asymmetry of the Universe (BAU)~\cite{Planck:2018vyg, Fields:2019pfx}. 
New physics beyond the SM is necessary to explain these phenomena. 

There are roughly two directions for such new physics models. 
Models along with the first direction are those which strongly depend on physics at very high energies, like models with the seesaw mechanism~\cite{ref: Type-I seesaw} and leptogenesis~\cite{Fukugita:1986hr}. 
On the other hand, models along with the second direction have a strong connection with the physics at TeV scales, for example, radiative seesaw models~\cite{Zee:1980ai, Cheng:1980qt, Zee:1985id, Babu:1988ki}, weakly interacting massive particles (WIMPs) as DM and electroweak baryogenesis (EWBG)~\cite{Kuzmin:1985mm}. 
Although both directions are important, it would be good timing to seriously consider the models with the second direction from the viewpoint of the testability at near future experiments. 
The Higgs boson was found in 2012~\cite{Higgs_discovery}. However, so little is known about the structure of the Higgs sector and the nature of electroweak symmetry breaking. 
Models along with the second direction are expected to be strongly related to physics of the non-standard Higgs sectors. 

In the radiative seesaw models, neutrino masses are generated by the quantum effects of additional scalar bosons~\cite{Zee:1980ai, Cheng:1980qt, Zee:1985id, Babu:1988ki}. 
Thus, an extended Higgs sector at the TeV scale plays an essential role. 
In addition, if the extended Higgs sector contains a new stable particle due to an additionally imposed symmetry, like $Z_2$ parity, it can be the dark matter candidate as the WIMP. 
Therefore, such models can explain dark matter simultaneously~\cite{Krauss:2002px}. 
Various radiative seesaw models along this line have been proposed so far~\cite{Krauss:2002px, Ma:2006km, ref: AKS, Gustafsson:2012vj, ref: Dirac1-loop, Gu:2007ug, Kanemura:2017haa, Enomoto:2019mzl}.

In Ref.~\cite{ref: AKS} (denoted by AKS09 in the following), a radiative seesaw model was proposed. In this model, neutrino mass, DM and BAU can be simultaneously explained by the new physics at the TeV scale without assuming an unnatural hierarchy among the mass scales. 
The Higgs sector of the model contains two Higgs doublet fields $\Phi_1$ and $\Phi_2$, a couple of charged singlet fields $S^\pm$ and a real singlet field $\eta$. 
The softly broken $Z_2$ symmetry is assumed for avoiding flavor-changing neutral currents (FCNCs) at tree level~\cite{Glashow:1976nt}. 
The Higgs doublets have the Type-X Yukawa interaction~\cite{Barger:1989fj, Grossman:1994jb, Aoki:2009ha, ref: LeptophilicTHDM}. 
In this model, three right-handed neutrinos $N_R^\alpha$ ($\alpha = 1,2,3$) are also added, which are odd under an unbroken $Z_2$ symmetry. 
$S^\pm$ and $\eta$ are also odd. 
Majorana masses of $N_R^\alpha$ violate lepton number conservation. 
Due to the unbroken $Z_2$ symmetry, it is prohibited to produce Majorana neutrino masses at tree level like the type-I seesaw mechanism~\cite{ref: Type-I seesaw}.  
Instead, they are generated at three-loop level.
The unbroken $Z_2$ symmetry also has an important role in DM physics. 
The lightest $Z_2$-odd particle is a DM candidate if it is electromagnetically neutral. Thus, $\eta$ or the lightest of $N^\alpha_R$'s can be the DM particle. 
The extra Higgs doublet is required not only for neutrino mass but also for EWBG. 
This model can provide new CP-violating phases, 
and at the same time, it can realize the strongly first-order electroweak phase transition (EWPT), 
which are required for successful EWBG. 
However, in AKS09, the authors have focused on the first-order EWPT and have not evaluated the baryon number generation neglecting CP-violating phases for simplicity. 
Therefore, it has been required to establish the complete model including the CP-violating phases and to propose a benchmark scenario that can simultaneously explain neutrino mass, DM and BAU with satisfying all the currently available data. 

In this paper, we evaluate BAU in an extended model of the original model proposed in AKS09. 
We do not impose the softly broken $Z_2$ symmetry in the original model and instead impose a flavor alignment structure of the Yukawa interactions to avoid FCNCs~\cite{Pich:2009sp}. 
Then, multiple CP-violating phases can be introduced in the Yukawa interaction as well as the Higgs potential. 
Due to their destructive interference, large CP-violating phases required for successful EWBG can be compatible with the severe constraint from the electric dipole moment (EDM) measurements~\cite{Kanemura:2020ibp, Kanemura:2021atq, Enomoto:2021dkl, Enomoto:2022rrl}. 
We discuss theoretical and experimental constraints on this model. 
Neutrino mass, lepton flavor violating processes and DM physics are also discussed, and some relevant formulae are shown.
We then evaluate the baryon number generation along the scenario of EWBG. 
The EWPT in the model is first discussed. 
Next, we describe how to evaluate the baryon number density by using the WKB method~\cite{Joyce:1994fu, Cline:2000nw, Fromme:2006wx, Cline:2020jre} in the model.
Consequently, we find a benchmark scenario that can explain neutrino mass, DM and BAU simultaneously with satisfying all the other available experimental data such as flavor observables, collider signals and EDM measurements.  
The benchmark scenario has rich phenomenological predictions which can be tested in various current and future experiments. 
The phenomenological impact of the model is also discussed.  

This paper is organized as follows. 
In the next section, we show the particle content of the model and a part of Lagrangian. 
The theoretical and experimental constraints on the model are also discussed in this section. 
In Sec.~\ref{sec: BSMphenomena}, formulae for neutrino mass, lepton flavor violating processes and DM physics are presented. 
In Sec.~\ref{sec: EWBG}, EWBG in the model is discussed. We first discuss EWPT in the model. Next, the mechanism of the baryon number generation in the model is explained.  
In Sec.~\ref{sec: Benchmark}, we show the benchmark scenario and some numerical evaluations. The prediction of the benchmark scenario is also discussed. 
Some discussions and conclusions are presented in Sec.~\ref{sec: conclusions}. 
In Appendices, we show some formulas omitted in the text and detailed derivations of some formulas in the text.

\section{Model}
\label{sec: Model}

In this section, Lagrangian of the model is given with the notation following AKS09.  
We first show most general Lagrangian.  
Then, we simplify Lagrangian with a few phenomenological assumptions for satisfying experimental constraints.
Theoretical and experimental constraints on the model are also discussed. 

\subsection{Lagrangian}
 
Fields in the model are shown in Table~\ref{table: contents}. 
In the model, a new unbroken $Z_2$ symmetry is imposed. 
Two Higgs doublets ($\Phi_1$ and $\Phi_2$) and the other SM particles are even under the $Z_2$ parity. 
On the other hand, the charged scalar bosons $S^\pm$, the real scalar boson $\eta$ and three Majorana fermions $N_{R}^\alpha$ ($\alpha=1,2,3$) are odd under the $Z_2$ parity. 
The Majorana fermions $N_R^\alpha$ have Majorana masses $m_{N^\alpha}^{}$. 
In AKS09, a softly broken $Z_2$ symmetry was also introduced, while we here do not impose this symmetry. 

The electroweak symmetry is broken by the vacuum expectation values (VEVs) of the Higgs doublets. 
In the following, we work on the Higgs basis~\cite{Davidson:2005cw}, in which only the neutral element of $\Phi_1$ obtains the real VEV without loss of generality. 

\begin{table}[t]
\begin{center}
\begin{tabular}{|c|cc|cc||ccccc|c|}  \hline
 & $\Phi_1$ & $\Phi_2$ & $S^+$ & $\eta$ & $Q_{L}^{\prime i}$ & $u^{\prime i}_{R}$ & $d_{R}^{\prime i}$ & $L_{L}^{\prime i}$ & $\ell_{R}^{\prime i}$ & $N_{R}^\alpha$ \\ \hline
 Spin & 0 & 0 & 0 & 0 & 1/2 & 1/2 & 1/2 & 1/2 & 1/2 & 1/2  \\
 $\mathrm{SU(3)_C}$ & {\bf 1} & {\bf 1} & {\bf 1} & {\bf 1} & {\bf 3} & {\bf 3} & {\bf 3} & {\bf 1} & {\bf 1} & {\bf 1} \\
 $\mathrm{SU(2)_L}$ & {\bf 2} & {\bf 2} & {\bf 1} & {\bf 1} & {\bf 2} & {\bf 1} & {\bf 1} & {\bf 2} & {\bf 1} & {\bf 1} \\
 $\mathrm{U(1)_Y}$ & $1/2$ & $1/2$ & $1$ & $0$ & $1/6$ & $2/3$ & $-1/3$ & $-1/2$ & $-1$ & $0$ \\ 
 $Z_2$ & $+$ & $+$ & $-$ & $-$ & $+$ & $+$ & $+$ & $+$ & $+$ & $-$ \\ \hline
 \end{tabular}
 \caption{Fields in the model. The indices $i$ and $\alpha$ represent the generation of the fermions. $\eta$ is a real scalar field.}
 \label{table: contents}
 \end{center}
 \end{table}

The Higgs potential is given by
\bal
\label{eq:potential_Higgsbasis}
V & = \sum_{a=1}^2 \Bigl( \mu_a^{2} |\Phi_a|^2 + \frac{ \lambda_a }{ 2 } |\Phi_a |^4 \Bigr)
+ \bigl( \mu_{12}^{2} \Phi_1^\dagger \Phi_2 + \mathrm{h.c.} \bigr)
+ \lambda_3 | \Phi_1 |^2 | \Phi_2 |^2 
+ \lambda_4 | \Phi_1^\dagger \Phi_2 |^2 
\nonumber \\
& + \Bigl\{ 
	\Bigl( \frac{ \lambda_5 }{ 2 } ( \Phi_1^\dagger \Phi_2 ) 
	+ \lambda_6 |\Phi_1|^2
	+ \lambda_7 | \Phi_2 |^2 
	\Bigr) (\Phi_1^\dagger \Phi_2) 
	+ \mathrm{h.c.}
	\Bigr\}
+ \mu_S^{2} | S^+ |^2 + \frac{ \mu_\eta^{2} }{ 2 } \eta^2 
\nonumber \\
&+ \Bigl\{ 
	\Bigl( \rho_{12} | S^+ |^2 + \frac{ \sigma_{12} }{ 2 } \eta^2 \Bigr) 
	(\Phi_1^\dagger \Phi_2 ) 
	+ 2 \kappa (\tilde{\Phi}_1^\dagger \Phi_2 ) S^- \eta + \mathrm{h.c.}
	 \Bigr\}
\nonumber \\
& + \sum_{a=1}^2 
	\Bigl( \rho_a | S^+ |^2 + \frac{ \sigma_a }{ 2 }\eta^2   \Bigr)
	| \Phi_a |^2 
+ \frac{ \lambda_S^{} }{ 4 } | S^+ |^4 
+ \frac{ \lambda_\eta }{ 4! } \eta^4 
+ \frac{ \xi }{ 2 } | S^+ |^2 \eta^2, 
\eal 
where $\mu_{12}^2$, $\lambda_5$, $\lambda_6$, $\lambda_7$, $\rho_{12}$, $\sigma_{12}$ and $\kappa$ are complex. 
The phase of $\kappa$ can be removed by rephasing $S^\pm$. 
One of $\mu_{12}^2$, $\lambda_5$, $\lambda_6$, $\lambda_7$, $\rho_{12}$ and $\sigma_{12}$ can be real by rephasing $\Phi_2$. 
Therefore, there are five CP-violating parameters in the Higgs potential. 

We define components of the Higgs doublets as 
\beq
\label{eq: Higgs doublets}
\Phi_1 = 
\begin{pmatrix}
G^+ \\
(v + H_1 + i G^0 )/ \sqrt{2} \\
\end{pmatrix}
, \quad 
\Phi_2 = 
\begin{pmatrix}
H^+ \\
( H_2 + iH_3 )/\sqrt{2} \\
\end{pmatrix}, 
\eeq
where $v \simeq 246~\mathrm{GeV}$ is the VEV of $\Phi_1$.  
The stationary condition of the vacuum requires
\beq
2 \mu_1^2 = -  \lambda_1 v^2, 
\quad
2 \mu_{12}^2 = - \lambda_6 v^2.
\eeq
The phase of $\mu_{12}^2$ is related to that of $\lambda_6$. 
Therefore, the Higgs potential includes four independent CP-violating phases.

By using Eq.~(\ref{eq: Higgs doublets}) and the stationary condition, 
masses of $H^\pm$, $S^\pm$ and $\eta$ are given by
\beq
m_{H^\pm}^2 = \mu_2^2 + \frac{ \lambda_3 }{ 2 } v^2, \quad 
m_S^2 = \mu_S^2 + \frac{ \rho_1 }{ 2 } v^2, \quad  
m_\eta^2 = \mu_\eta^2 + \frac{ \sigma_1 }{ 2 } v^2, 
\eeq
respectively. $G^{\pm}$ and $G^0$ in $\Phi_1$ are massless and Nambu--Goldstone (NG) bosons. 

$H_1$, $H_2$ and $H_3$ are not mass eigenstates at this stage. 
Their mass matrix $M_h^2$ is given by
\beq
\label{eq: mass_matrix_neutral_scalars}
M_h^2 = 
\begin{pmatrix}
\lambda_1 v^2 & \mathrm{Re}[\lambda_6] v^2 & - \mathrm{Im}[\lambda_6] v^2 \\[3pt]
\mathrm{Re}[\lambda_6] v^2 & M_+^2 & - \mathrm{Im}[\lambda_5]v^2 /2 \\[3pt]
- \mathrm{Im}[\lambda_6]v^2 & - \mathrm{Im}[\lambda_5]v^2 /2 & M_-^2 \\
\end{pmatrix}, 
\eeq
where $M_\pm^2 = \mu_2^2 + (\lambda_3 + \lambda_4 \pm \mathrm{Re}[\lambda_5])v^2 /2$. 
As mentioned above, $\mathrm{Im}[\lambda_5]$ can be zero by rephasing $\Phi_2$. 
Then, the non-diagonal terms are generated by only $\lambda_6$. 
In the following, we consider this case. 
Further discussion about mass eigenstates of neutral scalar states is given in the next section after imposing a phenomenological assumption for simplicity.

Next, we consider Yukawa interactions in the model. 
Both the Higgs doublets have Yukawa interactions with the SM fermions. 
\beq
\begin{array}{l}
- \mathcal{L}_Y = 
(y_u^a)_{ij} \overline{Q_{L}^{\prime i}}\tilde{\Phi}_a u_{R}^{\prime j} 
	 + (y_d^a)_{ij} \overline{Q_{L}^{\prime i} }\Phi_a d_{R}^{\prime j} 
	 + (y_\ell^a)_{ij} \overline{L_{L}^{\prime i} } \Phi_a \ell_{R}^{\prime j} + \mathrm{h.c.}, \\
\end{array}
\eeq
where the repeated indices are implicitly summed over. 
$Z_2$-odd fields $S^\pm$ and $N_R^\alpha$ also have Yukawa interactions;
\beq
\label{eq: YukawalSN_before}
- \mathcal{L}_{\ell S N} 
= h_i^{\prime \alpha} \overline{(N_{R}^\alpha)^c} \ell_{R}^{\prime i} S^+ + \mathrm{h.c.}
\eeq
The matrix $h^\prime$ has nine complex phases. 
Three of them can be zero by rephasing lepton fields without changing other terms in Lagrangian. 
Therefore, the matrix $h$ includes six CP-violating phases. 
This degree of freedom is the same as that for removing unphysical phases from the Pontecorvo--Maki--Nakagawa--Sakata (PMNS) matrix~\cite{ref: Pontecorvo, Maki:1962mu}. 

\subsection{Alignment scenario}

While the above discussion is general, we impose a few phenomenological assumptions in the following for simplicity. 
The non-diagonal terms of $M_h^2$ in Eq.~(\ref{eq: mass_matrix_neutral_scalars}), which are proportional to $\lambda_6$, induce mixings among $H_1$, $H_2$ and $H_3$. 
These mixings lead to the deviations of the $125~\mathrm{GeV}$ Higgs boson couplings from their SM predictions. Since such deviations are strictly constrained by the current LHC data~\cite{ref: HiggsCouplingLHC}, $\lambda_6$ is favored to be small. 
Thus, we assume that $\lambda_6 = 0$~\cite{Kanemura:2020ibp, Enomoto:2021dkl, Enomoto:2022rrl, Kanemura:2021dez, Kanemura:2021atq}. 
Then, $H_1$, $H_2$ and $H_3$ are mass eigenstates whose masses are given by
\beq
m_{H_1}^2 = \lambda_1 v^2, \quad m_{H_2}^2 = M_+^2, \quad m_{H_3}^2 = M_-^2. 
\eeq
$H_1$ corresponds to the SM-like Higgs boson at tree level.

The assumption $\lambda_6 = 0$ and the stationary condition lead to $\mu_{12}^2 = 0$. 
Since we set $\mathrm{Im}[\lambda_5]$ to be zero, $\lambda_7$, $\rho_{12}$ and $\sigma_{12}$ are CP-violating couplings in the Higgs potential. 
We define their phases as $\theta_7^{}$, $\theta_\rho^{}$ and $\theta_\sigma^{}$, respectively. 

Next, we discuss a simplification in the Yukawa interaction. 
In general, $y^1_f$ and $y^2_f$ ($f=u$, $d$, $\ell$) are not simultaneously diagonalized by the same biunitary transformation. 
This fact leads to the FCNCs at tree level~\cite{Glashow:1976nt}, which are strongly constrained by the current data of the flavor experiments. 
We avoid the dangerous FCNCs by simply assuming that $y_f^1$ and $y_f^2$ can be simultaneously diagonalized, i.e. $y^1_f$ and $y^2_f$ are assumed to be transformed as follows by a single biunitary transformation; 
\begin{equation}
\label{eq: biunitary_tr}
(y_f^1)_{ij} \to \left(\frac{ m_{f^i}^{} }{ v } \right)\delta_{ij},  \quad 
(y_f^2)_{ij} \to \zeta_{f^i}^{} \left(\frac{ m_{f^i}^{} }{ v }\right) \delta_{ij}, 
\end{equation}
where $m_{f^i}^{}$ are masses of the SM fermions, and $\zeta_{f^i}^{}$ are complex parameters.
In Eq.~(\ref{eq: biunitary_tr}), the summation for the same indices is not implemented.
Then, the strength of the Yukawa interactions between $\Phi_2$ and the SM fermions is controlled by $\zeta_{f^i}$.

In addition, we make further simplifications for the quark Yukawa couplings. 
We assume the flavor universality for $\zeta_{u^i}$ and  $\zeta_{d^i}$. 
The Yukawa interactions between $\Phi_2$ and quarks are then described by two parameters $\zeta_u$ and $\zeta_d$. 
It is the same situation in two Higgs doublet models (THDMs) with Yukawa alignment~\cite{Pich:2009sp}. 
On the other hand, we allow the flavor dependence for the lepton Yukawa couplings $(\zeta_{\ell^1}, \zeta_{\ell^2}, \zeta_{\ell^3}) = (\zeta_e, \zeta_\mu, \zeta_\tau)$ so as to explain neutrino mass data as discussed later.  

By using the above assumptions for the simplification, 
the Yukawa interactions between the SM fermions and the Higgs bosons are given by 
\bal
\label{eq: Higgs_Yukawa}
-\mathcal{L}_Y = & \frac{ \sqrt{2} }{ v }
\biggl\{
	\overline{u^i} V_{ij} ( \zeta_d m_{d^j}^{} P_R - \zeta_u m_{u^i}^{} P_L ) d^jH^+ 
	+ \zeta_{\ell^i} m_{\ell^i}^{}  \overline{\nu^{i}} P_R \ell^i H^+ 
	+ \mathrm{h.c.}
\biggr\} 
\nonumber \\
& +  \frac{ m_{f^i} }{ v }
\overline{f^i} (Z^k_{f^i} + i X^k_{f^i} \gamma_5) f^i H_k + \cdots, 
\eal
where $\cdots$ denotes the other terms in $\mathcal{L}_Y$. 
The fermions without ${}^\prime$ represent the mass eigenstates. 
The matrix $V$ is the Cabibbo--Kobayashi--Maskawa matrix~\cite{Cabibbo:1963yz, Kobayashi:1973fv}. 
The coefficients $Z^k_{f^i}$ and $X^k_{f^i}$ ($k=1$, 2, 3) are defined as 
\bal
\label{eq: Z_Xfactor}
& \left\{
\begin{array}{lll}
Z_{u^i}^1 = 1, & Z_{d^i}^1 = 1, & Z_{\ell^i}^1 = 1, \\
Z_{u^i}^2 = \mathrm{Re}[\zeta_u], & Z_{d^i}^2 = \mathrm{Re}[\zeta_d], & Z_{\ell^i}^2 = \mathrm{Re}[\zeta_{\ell^i}], \\
Z_{u^i}^3 = - \mathrm{Im}[\zeta_u], & Z_{d^i}^3 = - \mathrm{Im}[\zeta_d], & Z_{\ell^i}^3 = -\mathrm{Im}[\zeta_{\ell^i}], 
\end{array} \right. \\
& \left\{
\begin{array}{lll}
X_{u^i}^1 = 0, & X_{d^i}^1 = 0, & X_{\ell^i}^1 = 0, \\
X_{u^i}^2 = - \mathrm{Im}[\zeta_u], & X_{d^i}^2 = \mathrm{Im}[\zeta_d], & X_{\ell^i}^2 = \mathrm{Im}[\zeta_{\ell^i}], \\
X_{u^i}^3 = - \mathrm{Re}[\zeta_u], & X_{d^i}^3 = \mathrm{Re}[\zeta_d], & X_{\ell^i}^3 = \mathrm{Re}[\zeta_{\ell^i}]. 
\end{array}
\right. 
\eal

After the biunitary transformation, the Yukawa interactions among $S^\pm$, $N_R^\alpha$ and the charged leptons are given by 
\beq
\label{eq: YukawalSN}
- \mathcal{L}_{\ell S N} 
= h_i^{\alpha} \overline{(N_{R}^\alpha)^c} \ell_{R}^{i} S^+ + \mathrm{h.c.}, 
\eeq
where the coupling constants $h_i^\alpha$ are defined in the mass eigenstate basis of the charged leptons.

In this section, we have assumed the alignment structures in the Higgs potential and the Yukawa interactions for simplicity. 
Considering radiative corrections, these assumptions are broken. 
However, their effects are expected to be small and would not largely change the following discussion~\cite{Kanemura:2020ibp}.

\subsection{Theoretical constraints}
\label{sec: theoretical_constraints}

We here discuss theoretical constraints on the model from the vacuum stability, the triviality bound and the perturbative unitarity. 
The vacuum stability in the original model has been studied in Ref.~\cite{Aoki:2011zg}. They required that the Higgs potential is bounded from below in any direction of the scalar fields and found eleven inequalities of the scalar couplings below the cut-off scale $\Lambda$. We use their result. See Ref.~\cite{Aoki:2011zg} for the explicit formulae of the inequalities. 

The triviality bound in the original model has also been discussed in Ref.~\cite{Aoki:2011zg}. They investigated constraints so that all the running scalar couplings do not blow up nor fall down below $\Lambda$. 
Referring to their results, it is simply supposed that absolute values of all the dimensionless scalar couplings are less than two in order to keep $\Lambda$ to be larger than $10~\mathrm{TeV}$. 

The perturbative unitarity requires that 
each wave amplitude $a_\ell^{}$ is bounded within a circle in the complex plane described by $|2 a_\ell^{} - i| = 1$~\cite{Lee:1977eg}. 
At tree level, as the constraint for the Higgs potential, 
it is often imposed that $|a_0^{}| < 1/2$ for all two-to-two-body scalar boson scatterings.
This constraint can be generally satisfied as long as the scalar couplings are not too large, or equivalently, the masses of the additional Higgs bosons ($H_2$, $H_3$ and $H^\pm$), $S^\pm$ and $\eta$ are not far from $\mu_2^2$, $\mu_S^2$ and $\mu_\eta^2$, respectively~\cite{Kanemura:1993hm, Akeroyd:2000wc, Ginzburg:2005dt, Kanemura:2015ska}. 
We now consider the scalar couplings less than two as a criterion to satisfy the above triviality bound. 
We assume that this is also sufficient to satisfy the perturbative unitarity. 
More detailed analyses on the above theoretical constraints will be performed elsewhere~\cite{ref: futureAKS}. 

\subsection{Experimental constraints}
\label{sec: experimental_constraints}

We here discuss constraints on the model from current experimental data. 
For simplicity, we focus on the case that $m_{N^\alpha} = \mathcal{O}(1)~\mathrm{TeV}$ and masses of the other new particles are around the electroweak scale. $\eta$ is then the DM particle. 
This setup is valid in the benchmark scenario discussed in Sec.~\ref{sec: Benchmark}. 

First, we give a quick review of direct searches for the additional Higgs bosons at high-energy colliders. 
We use the theoretical results in Refs.~\cite{Arbey:2017gmh, Aiko:2020ksl} and experimental data from LEP and LHC.  
By the direct search for $H^\pm$ at LEP, 
the lower bound on $m_{H^\pm}$ is given by $m_{H^\pm} \gtrsim 80~\mathrm{GeV}$~\cite{ALEPH:2013htx}. 
This bound is almost independent of the value of $|\zeta_{f^i}^{}|$. 
At LHC, light $H^\pm$ can be produced via the decay of top quarks and predominantly decays into $\tau \nu$. 
The latest result of the search for this signal is given in Ref.~\cite{ref: lightHpmSearch}. 
When all $\zeta_{f^i}^{}$ couplings are real and have the same absolute value, 
the upper bound on them is given by $|\zeta_{f^i}^{}| \lesssim 0.06$ ($0.04$) for $m_{H^\pm} = 100~\mathrm{GeV}$ ($150~\mathrm{GeV}$).  
If $m_{H^\pm} > m_t - m_b \simeq 170~\mathrm{GeV}$, $H^\pm$ is predominantly produced via the associated production; $gg \to tbH^\pm$. 
Then, the main decay mode of $H^\pm$ is $tb$ unless $|\zeta_u|$ is too small. 
The latest search for this signal is given in Ref.~\cite{ATLAS:2021upq}. 
In the case that all $\zeta_{f^i}^{}$ have the same value, the upper limit for $|\zeta_{f^i}^{}|$ is given by $|\zeta_{f^i}^{}| \lesssim 0.5$ ($0.6$) for $m_{H^\pm} = 200~\mathrm{GeV}$ ($400~\mathrm{GeV}$).  

The additional neutral Higgs bosons $H_2$ and $H_3$ are predominantly generated via the gluon-fusion process at LHC.\footnote{The bottom quark associated production is also important in the case that $|\zeta_d|$ is much larger than $|\zeta_u|$ like the MSSM with a large $\tan \beta$. }
If $H_2$ and $H_3$ are lighter than $2 m_t \simeq 350~\mathrm{GeV}$, 
they decay into $\tau \overline{\tau}$ or $b\overline{b}$. 
The latest result of the search for $H_{2,3} \to \tau \overline{\tau}$ is given in Ref.~\cite{ATLAS:2020zms}. 
The upper bound on $|\zeta_{f^i}^{}|$ is estimated as 0.35 for $m_{H_{2,3}} = 200~\mathrm{GeV}$. 
Heavier $H_{2,3}$ can decay into $t \bar{t}$. 
Given the same real-valued $|\zeta_{f^i}^{}|$, the constraint from $H_{2,3}^{} \to t \bar{t}$ is given by $|\zeta_{f^i}^{}| \lesssim 0.67$ for $m_{H_{2,3}} = 400~\mathrm{GeV}$ according to the latest data in Ref.~\cite{ATLAS:2018rvc}. 

The additional Higgs bosons are also produced by pair productions via electroweak gauge bosons~\cite{Kanemura:2001hz, Cao:2003tr, Belyaev:2006rf, Kanemura:2011kx, Kanemura:2021dez}.
In particular, in Ref.~\cite{Kanemura:2021dez}, it has been investigated the pair productions in the Higgs alignment scenario of the THDM with the flavor-aligned Yukawa interaction like the present model but neglecting the CP-violating phases. 
It has been found that the search for the multilepton final states at LHC strongly constrains the additional Higgs bosons whose masses are smaller than $2m_t$ in the case that $|\zeta_{\ell^i}|$ is much larger than $|\zeta_d|$. 
On the other hand, in the case that $|\zeta_d|\simeq|\zeta_{\ell^i}^{}|$, 
there is almost no constraint from the multilepton search because the additional Higgs bosons predominantly decay into a pair of quarks. 
This constraint is also relaxed for the heavier additional Higgs bosons than $2m_t$ because $H_{2,3} \to t \overline{t}$ is kinematically allowed. 

A pair of $S^\pm$ is produced via the electroweak gauge bosons $Z$ and $\gamma$ at hadron colliders. 
At $e^+ e^-$ colliders, there is an additional t-channel diagram for the pair production via the Yukawa couplings $h_i^\alpha$~\cite{ref: AKS}.
Since we consider the case of $m_{N^\alpha}^{} > m_S$, 
$S^\pm$ predominantly decay into $H^{\pm} \eta$ via the $\kappa$ coupling in the Higgs potential.  
$H^{\pm}$ predominantly decay into $\tau \nu$ or $tb$ depending on their mass. 
For lighter $H^\pm$, the main signal from $S^+S^-$ pair production is $\tau^- \tau^+ \cancel{E}$, where $\eta$ and $\nu_\tau$ observed as a missing energy. 
Such a final state is well investigated in the context of the stau searches in the supersymmetric models and strongly constrained from the latest LHC data~\cite{ATLAS:2019gti}. 
On the other hand, the main decay mode of $S^\pm$ in the case of heavier $H^\pm$ is $t \bar{t} b \bar{b} \cancel{E}$.
The constraint on such a signal is expected to be weak enough at both $e^+e^-$ colliders and hadron colliders. 

A pair of $N^\alpha_R$'s are produced at $e^+e^-$ colliders via t-channel and u-channel diagrams~\cite{ref: AKS}. 
Since $m_{N^\alpha}$ is at a few TeV, the production cross section is so small that no effective constraint is obtained from the current data. 
At hadron colliders, $N^\alpha_R$ productions are loop-induced processes or much suppressed tree-level processes because $N^\alpha_R$ does not have couplings with quarks at tree level. 
The current LHC data would thus give almost no constraint on $N^\alpha$'s. 

Next, we discuss constraints from the flavor experiments. 
The charged Higgs bosons are constrained by $B_d^0 \to \mu^+ \mu^-$, $B \to X_s \gamma$, $B^0$-$\overline{B^0}$, and so on~\cite{Arbey:2017gmh, Enomoto:2015wbn, Haller:2018nnx}. 
In the case that all $|\zeta_{f^i}^{}|$ have the same value, the strongest constraint comes from $B_d^0 \to \mu^+ \mu^-$. 
The upper limit on $|\zeta_{f^i}|$ is given by $|\zeta_{f^i}| \lesssim 0.33$ ($|\zeta_{f^i}| \lesssim 0.5$) for $m_{H^\pm} = 100~\mathrm{GeV}$ ($m_{H^\pm} = 400~\mathrm{GeV}$) at $95~\%$ C.L. 

The Yukawa interaction $\mathcal{L}_{\ell SN}$ in Eq.~(\ref{eq: YukawalSN}) induces the lepton flavor violating processes. 
Various experiments search for such processes and give so strong constraints on new physics violating lepton flavor conservation~\cite{MEG:2016leq, BaBar:2009hkt, SINDRUM:1987nra, Hayasaka:2010np}.  
This constraint in the model is discussed in detail in Sec.~\ref{sec: BSMphenomena} and~\ref{sec: Benchmark}. 

Finally, we discuss constraints on the CP-violating phases in the model. 
The strongest constraint on the CP-violation is given by the electron EDM (eEDM) measurements~\cite{ACME:2018yjb, Roussy:2022cmp}. 
The current upper limit for the eEDM is given by $|d_e| < 4.1 \times 10^{-30}~\mathrm{e\, cm}$ with $90~\%$ C.L.~\cite{Roussy:2022cmp}.
In the model, the leading contribution to the eEDM comes from the Barr-Zee type diagrams~\cite{Barr:1990vd}. 
The diagrams including the loop of the additional Higgs bosons and $S^\pm$ are proportional to $\sin (\theta_7 - \theta_e^{})$ and $\sin(\theta_\rho^{} - \theta_e^{})$, respectively, where $\theta_e^{}$ is the phase of $\zeta_e^{}$~\cite{Kanemura:2020ibp}.
On the other hand, the diagram including the top quark loop is proportional to $\sin (\theta_u^{} - \theta_e^{})$. 
In the case that $|\zeta_u| = |\zeta_d| = |\zeta_{\ell^i}|$, 
the loop diagrams of other fermions are negligibly small. 
Thus, three independent phases $\theta_7^{} - \theta_e^{}$, $\theta_\rho^{} - \theta_e^{}$ and $\theta_u^{} - \theta_e^{}$ mainly contribute to the eEDM in the model. 
By using the destructive interferences among these phases, the eEDM can be smaller than the current upper bound~\cite{Kanemura:2020ibp, Kanemura:2021atq, Enomoto:2021dkl, Enomoto:2022rrl}. 
Other EDM measurements, for example, the neutron EDM (nEDM), also give constraints on the model~\cite{Abel:2020pzs}. 
However, we do not consider them because they are weaker than the constraint from the eEDM in the parameter region discussed below~\cite{Kanemura:2020ibp, Kanemura:2021atq, Enomoto:2021dkl, Enomoto:2022rrl}.  

\section{Neutrino mass, lepton flavor violation and dark matter}
\label{sec: BSMphenomena}

In this section, we discuss neutrino mass, lepton flavor violation and DM in the model. 
Some formulae for them are presented. 

\subsection{Neutrino mass}
\label{sec: neutrino mass}

\begin{figure}[t]
\begin{center}
\includegraphics[width=0.6\textwidth]{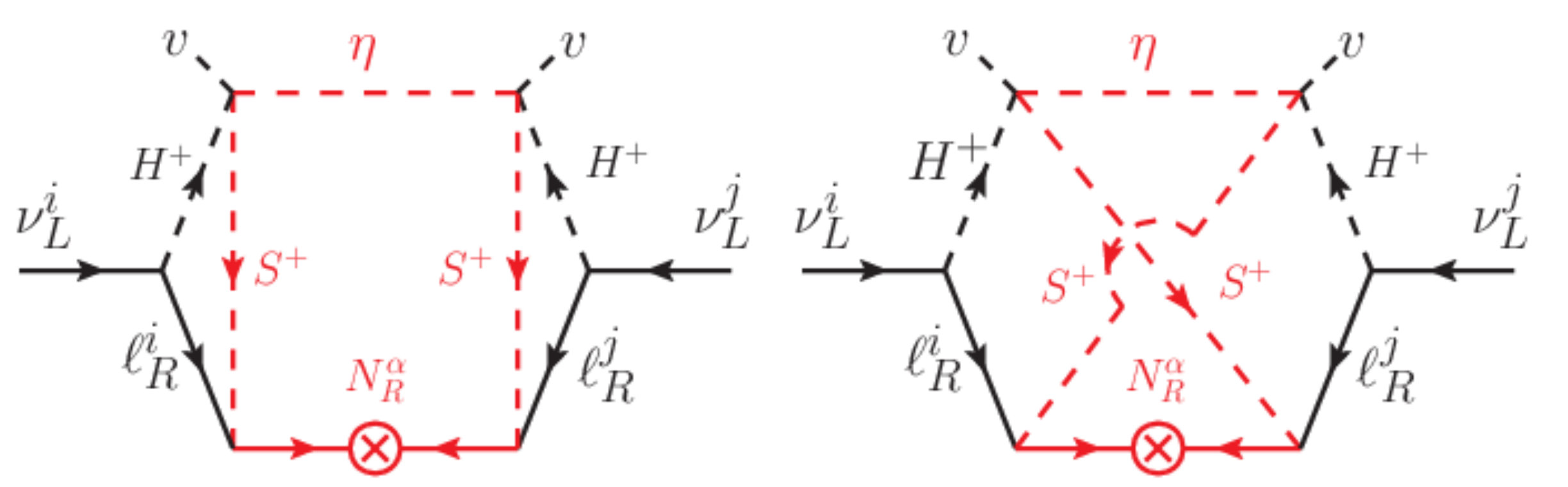}
\caption{Feynman diagrams for neutrino masses. Red lines are $Z_2$-odd fields. The symbol $\otimes$ means the Majorana masses of $N_R^\alpha$.}
\label{fig: Neutrino_mass_diagram}
\end{center}
\end{figure}

In this model, Majorana neutrino masses are generated at three-loop level. 
The Feynman diagrams are shown in Fig.~\ref{fig: Neutrino_mass_diagram}~\cite{ref: AKS}. 
The neutrino mass matrix $M_\nu$ is evaluated as
\beq
\label{eq: formula_neutrino_mass}
(M_{\nu})_{ij}
=  \frac{ \kappa^2 (\zeta_{\ell^i}^\ast m_{\ell^i} ) (\zeta_{\ell^j}^\ast m_{\ell^j})  }{ (16 \pi^2 )^3} 
\sum_{\alpha = 1}^3 h^\alpha_i h^\alpha_j m_{N^\alpha}
\Bigl( F_{1\alpha} + F_{2\alpha} \Bigr). 
\eeq
The loop functions $F_{1\alpha}$ and $F_{2\alpha}$ correspond to the left and right diagrams in Fig.~\ref{fig: Neutrino_mass_diagram}, respectively, which are given by
\beq
\label{eq: formula_Fna}
F_{n\alpha} = \int_0^1 \tilde{\mathrm{d}}^4 x 
\int_0^\infty \mathrm{d}u \int_0^\infty \mathrm{d}v
\frac{  8 \sqrt{uv} \tilde{F}(a_{n}^{}, b_n^{}) }{ (u + m_{H^\pm}^2)(v + m_{H^\pm}^2) },  
\eeq
where
\beq
\int_0^1 \tilde{ \mathrm{d} }^4 x=\int_0^1 \mathrm{d}x \int_0^1\mathrm{d}y  \int_0^1 \mathrm{d}z  \int_0^1 \mathrm{d}w \,\delta (1-x-y-z-w).
\eeq
In Eq.~(\ref{eq: formula_neutrino_mass}), 
we consider only the leading terms of the charged lepton masses.
The function $\tilde{F}(x,y)$ and the arguments $a_n$ and $b_n$ are defined as follows;
\bal
& \tilde{F}(x, y) = \frac{ 1 }{ y^3 } \left( \sqrt{x^2 -y^2} + \frac{ x^2 }{ \sqrt{x^2 - y^2} } - 2 x \right), \\[5pt]
&  a_{1} = (y+z)m_S^2 + x m_{N^\alpha}^2 + \omega m_\eta^2 
	+ z (1-z)u + y(1-y)v, \\[5pt]
& a_{2} = (y+z) m_S^2 + x m_{N^\alpha}^2 + \omega m_\eta^2 
	+ (y+\omega)(x+z)u + (x+y)(z+\omega)v, \\[5pt]
& b_1 = 2 y z \sqrt{uv}, \quad b_2 = 2 (yz - x\omega) \sqrt{uv}. 
\eal
See Appendix~\ref{app: neutrino_formula} for the derivation of the neutrino mass matrix formula.

The loop function $F_{1\alpha}$ has a more simple integral formula~\cite{ref: AKS};
\bal
\label{eq: F1_simple_formula}
F_{1\alpha} = & \frac{ 4 }{ m_{H^\pm}^4 (m_{N^\alpha}^2 - m_\eta^2)} 
\int_0^\infty \mathrm{d}x\, x \left( \frac{ m_{N^\alpha}^2 }{ x + m_{N^\alpha}^2 } - \frac{ m_\eta^2 }{ x + m_\eta^2 } \right)
\nonumber \\
& \times \Bigl( B_1(m_{H^\pm}^2, m_S^2; -x) - B_1(0, m_S^2; -x) \Bigr)^2, 
\eal
where $B_1$ is the tensor coefficient in the formalism by Passarino-Veltman for one-loop integrals~\cite{Passarino:1978jh}. 
This simple representation is due to the topological property of the Feynman diagram. 
The left diagram in Fig.~\ref{fig: Neutrino_mass_diagram} can be divided into two one-loop diagrams by cutting the internal lines of $N_R^\alpha$ and $\eta$. 
The derivation of Eq.~(\ref{eq: F1_simple_formula}) is also shown in Appendix~\ref{app: neutrino_formula}.

Our formula is partly different from that in AKS09 even considering the extension explained in Sec.~\ref{sec: Model}. 
The function $F_{1\alpha}$ corresponding to the left figure in Fig.~\ref{fig: Neutrino_mass_diagram} is consistent. On the other hand, the formulae for $F_{2\alpha}$ disagree with each other. 
In AKS09, $F_{1\alpha} = F_{2\alpha}$, while we numerically found that $F_{2\alpha}$ in our formula gives only one order smaller contribution than $F_{1\alpha}$.

\subsection{Lepton flavor violatioin}
\label{sec: LFV}

The Yukawa couplings $h^\alpha_i$ in Eq.~(\ref{eq: YukawalSN}) induce lepton flavor violating processes as mentioned in Sec.~\ref{sec: experimental_constraints}. 
In particular, accessible processes in the model are $\ell^i \to \ell^j \gamma$ ($i \neq j$) and $\ell^i \to \ell^j \ell^k \overline{\ell^m}$ ($i \neq j$, $k$, $m$), which are generated at one-loop level~\cite{ref: AKS, Aoki:2011zg}. 
The branching ratio for $\ell^i \to \ell^j \gamma$ is given by~\cite{ref: AKS, Aoki:2011zg}
\beq
\label{eq: LFV1_formula}
\frac{ \mathrm{Br} ( \ell^i \to \ell^j \gamma ) }{ \mathrm{Br}(\ell^i \to \ell^j \nu_{\ell^i} \overline{\nu}_{\ell^j} ) }
= \frac{ 3 \alpha }{ 64 \pi G_F^2 }
\left| \sum_\alpha \frac{ (h^\alpha_j)^\ast h^\alpha_i }{ m_S^4 } f\Bigl( \frac{ m_{N^\alpha}^2 }{ m_S^2 } \Bigr) \right|^2, 
\eeq
where $\alpha$ and $G_F$ denote the fine structure constant and the Fermi coupling constant, respectively. 
The function $f(x)$ is defined as
\beq
f(x) = \frac{ 1 - 6 x + 3 x^2 + 2 x^3 - 6 x \log x }{ 6 (-1 + x )^4 }. 
\eeq

The three lepton decay processes $\ell^i \to \ell^j \ell^k \overline{\ell^m}$ are generated by photon penguin diagrams and box diagrams~\cite{Aoki:2011zg}. 
The box diagrams mainly contribute to the process with the large Yukawa coupling $h_i^\alpha$ that is favored to explain the neutrino oscillation data. 
Thus, we neglect the contribution of the photon penguin diagrams. 
The branching ratio is given by~\cite{Aoki:2011zg}
\beq
\label{eq: LFV2_formula}
\frac{ \mathrm{Br}( \ell^i \to \ell^j \ell^k \overline{\ell^m} ) }
{\mathrm{Br} (\ell^i \to \ell^j \nu_{\ell^i} \overline{\nu}_{\ell^j} )}%
=
\frac{ 2 - \delta_{jk} }{ 4096 \pi^2 G_F^2 }
\bigl| A_{i,m}^{jk}  \bigr|^2. 
\eeq
The tensor $A_{i,m}^{jk}$ is defined by using other two tensors $B_{i,m}^{jk}$ and $C_{i,m}^{jk}$ as 
\beq
A_{i,m}^{jk} = B_{i,m}^{jk} + B_{i,m}^{kj} + C_{i,m}^{jk} + C_{i,m}^{kj}. 
\eeq 
The tensors $B_{i,m}^{jk}$ and $C_{i,m}^{jk}$ are given by
\bal
& B_{i,m}^{jk} = 
	\frac{ h^\alpha_i (h^\alpha_j)^\ast (h^\beta_k)^\ast h^\beta_m }{ m_S^2 }
	\left(\frac{g_2\left(r_\alpha \right) - g_2\left(r_\beta\right) }{ r_\alpha - r_\beta } \right), \\[7pt]
& C_{i,m}^{jk} = 
	\frac{ h^\alpha_i (h^\beta_j)^\ast (h^\beta_k)^\ast h^\alpha_m }{ m_S^2 }
	\left( \frac{ g_1(r_\alpha) - g_1(r_\beta) }{ \sqrt{r_\alpha/r_\beta} - \sqrt{r_\beta/r_\alpha} } \right), 
\eal
where $r_\alpha = m_{N^\alpha}^2 / m_S^2$.
The function $g_k(x)$ ($k=1$, 2) is defined as
\beq
g_k(x) = \frac{ 1 }{ 2 } 
	\Bigl\{ 
		\frac{ x^k }{ (1 - x)^2 } \log x
		+ \frac{ 1 }{ 1 - x }
	\Bigr\}. 
\eeq

\subsection{Dark matter}
\label{sec: DM}

In the model, there are two kinds of $Z_2$-odd neutral particles, 
the real scalar boson $\eta$ and the Majorana fermions $N^\alpha_R$. 
The scalar boson $\eta$ or the lightest Majorana fermion can be the dark matter candidate. 
In the following, we consider the case that $\eta$ is the lightest $Z_2$-odd particle because heavy $N_R^\alpha$'s are favored to suppress the LFV processes. (See Eqs.~(\ref{eq: LFV1_formula}) and (\ref{eq: LFV2_formula}).)
We discuss the relic abundance of $\eta$ generated via the freeze-out mechanism.  

A pair of $\eta$ annihilates into a pair of fermions $f^i \bar{f^i}$ ($f=u$, $d$, $\ell$ and $i=1$, 2, 3), that of weak bosons $ZZ$ and $W^+W^-$ at tree level if it is kinematically allowed. 
At one-loop level, it also annihilates into a pair of photons~\cite {ref: AKS}. 
We do not include other channels, for example, annihilations into two scalar bosons because they are expected to contribute little in our benchmark scenario discussed in Sec.~\ref{sec: Benchmark}.

The thermally averaged annihilation cross section at the temperature $T$ is evaluated as~\cite{Gondolo:1990dk}
\beq
\left< \sigma v \right> = 
		\frac{ 2 x }{ K_2(x)^2 } 
		\int_1^\infty  \mathrm{d}y\  \sigma_{\text{\sf all}} vy \sqrt{y-1} K_1(2 x \sqrt{y}), 
\eeq
where $x = m_\eta / T$, $y = s/(4 m_\eta^2)$ with the Mandelstam variable $s$, $K_1$ and $K_2$ are the modified Bessel functions of the second kind,  
$v$ is the M\o lloer velocity~\cite{Gondolo:1990dk}, and $\sigma_{\text{\sf all}}$ is the sum of the annihilation cross section. 

The annihilation cross section for $\eta \eta \to f^i \bar{f^i}$ is given by 
\bal
\label{eq: fermionic_annihi}
(\sigma v)_{f^i}
= \frac{ N_c^f  m_{f^i}^2 }{ 4 \pi } 
\Biggl\{
	& h(m_{f^i})^3
	\left| \sum_{a=1}^3 \frac{ \lambda_{\eta \eta a } Z_{f_i}^a}{ s - m_{H_a}^2 + i m_{H_a}^{} \Gamma_{H_a} } \right|^2
	\nonumber \\[5pt]
	& + h(m_{f^i})  \left| \sum_{a=1}^3 \frac{ \lambda_{\eta \eta a } X_{f^i}^a}{ s - m_{H_a}^2 + i m_{H_a}^{} \Gamma_{H_a} } \right|^2
\Biggr\}, 
\eal
where $h(x) = \sqrt{1-4x^2/s}$, $\Gamma_{H_k}^{}$ is the decay width of $H_k$, $N_c^f$ is the color factor, and $\lambda_{\eta \eta \alpha}$ is given by 
\beq
\label{eq: EtaEtaH}
\lambda_{\eta \eta 1} = \sigma_1, \quad 
\lambda_{\eta \eta 2} = \mathrm{Re}[\sigma_{12}], \quad 
\lambda_{\eta \eta 3} = \mathrm{Im}[\sigma_{12}].
\eeq
We note that the CP-violating coupling $\lambda_{\eta \eta 3}$ induces the additional contribution to the annihilation cross section from the $H_3^{}$ mediated diagram compared with AKS09 neglecting the CP-violation. 

The annihilation cross section for $\eta \eta \to VV$ ($V = Z$ or $W^\pm$) is given by 
\beq
\label{eq: vector_annihi}
(\sigma v)_{V} = \frac{ S_V \sigma_1^2 }{ 16 \pi s } h(m_V^{})
\frac{ (s - 2 m_V^2 )^2 + 8 m_V^4 }{ (s - m_{H_1}^2)^2 + m_{H_1}^2 \Gamma_{H_1}^2 }, 
\eeq
where $S_V$ is one for the Z boson and two for the W boson. 
We note that there is no contribution from $H_2^{}$ and $H_3^{}$ at tree level because we assume the alignment in the Higgs potential ($\lambda_6 = 0$). 

\begin{figure}[t]
\begin{center}
\includegraphics[width=0.8\textwidth]{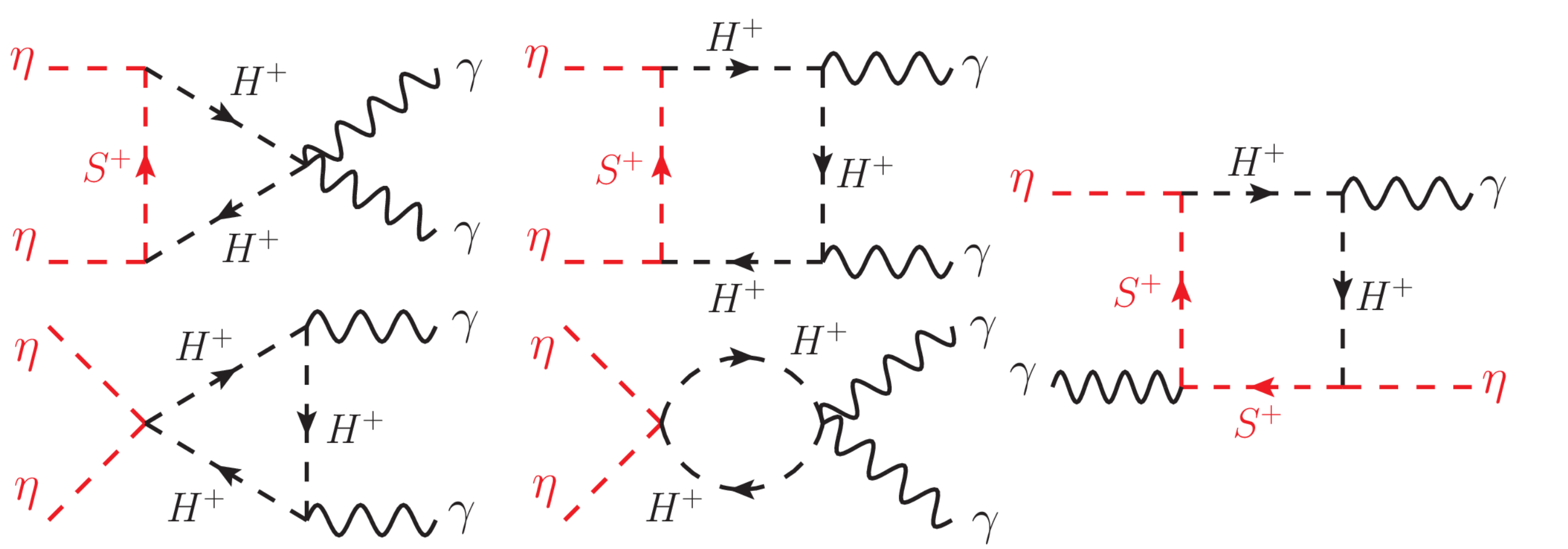} 
\caption{Feynman diagrams for $\eta \eta \to \gamma \gamma$. For the left four diagrams, there are also diagrams where $S^\pm$ and $H^\pm$ are interchanged. 
In addition, it includes diagrams with an appropriate momentum exchange in the initial and final states.}
\label{fig: diphoton}
\end{center}
\end{figure}

A pair of $\eta$ annihilates into two photons via one-loop diagrams including $S^\pm$ and $H^\pm$. The Feynman diagrams are shown in Fig.~\ref{fig: diphoton}.
In AKS09, the right box diagram in Fig.~\ref{fig: diphoton} is not included, and the result violates the Ward-Takahashi identity. We evaluated all the diagrams by performing the loop calculation. 
The annihilation cross section is given by 
\begin{align}
\label{eq: diphoton_annihi}
(\sigma v)_\gamma = \frac{ \alpha^2 }{ 512 \pi^3 s }
	\int_{-1}^1 \mathrm{d}\cos \theta\, 
	\Bigl(
		& 2 s^2 |A_{\gamma\gamma}|^2 
		+ 2 s^2 m_\eta^2 
			\bigl( A_{\gamma \gamma} B^\ast_{\gamma \gamma}
				+ A_{\gamma \gamma}^\ast B_{\gamma \gamma} \bigr)
			\nonumber \\
			& + 2 |B_{\gamma \gamma} |^2 
			\bigl( (m_\eta^4 - ut ) + s^2 m_\eta^4 \bigr)
	\Bigr), 
\end{align}
where $A_{\gamma \gamma}$ and $B_{\gamma \gamma}$ are loop functions depend on $s$ and the scattering angle $\cos \theta$. 
The explicit formula for $A_{\gamma \gamma}$ and $B_{\gamma \gamma}$ are shown in Appendix~\ref{app: 2Eta2gamma}.
In the case that $m_\eta, s \ll m_S, m_{H^\pm}$, 
$(\sigma v)_\gamma$ can be approximately evaluated by 
\beq
(\sigma v)_\gamma \simeq 
\frac{ s \alpha^2 }{ 96 \pi^3 }
	\biggl|
		\frac{ \kappa^2 v^2 }{ m_{H^\pm}^2 m_S^2 }
		- \frac{ 1 }{ 4 } 
		\Bigl( \frac{ \sigma_2 }{ m_{H^\pm}^2 }
			+ \frac{ \xi }{ m_S^2 } \Bigr)
	\biggr|^2. 
\eeq
We numerically checked that this formula gives a good approximation if $m_\eta$ and $\sqrt{s}$ are smaller than about half of $m_S^{}$ and $m_{H^\pm}^{}$. 

By using Eqs.~(\ref{eq: fermionic_annihi}), (\ref{eq: vector_annihi}) and (\ref{eq: diphoton_annihi}), we evaluate $\left< \sigma v \right>$. The relic abundance is then approximately given by~\cite{Gondolo:1990dk}
\bal
\label{eq: Relic_density}
\Omega_\eta h^2 \simeq 8.5 \times 10^{-11}
\biggl[
	\int_{T_0}^{T_f} g_\ast^{1/2} \left(\frac{\left< \sigma v \right>}{ \mathrm{GeV}^{-2} } \right)\frac{ \mathrm{d}T }{ m_\eta }
\biggr]^{-1}, 
\eal 
where $T_0 \simeq 2.73~\mathrm{K}$ is the present temperature of the cosmic microwave background~\cite{PDG2022}. 
See Ref.~\cite{Gondolo:1990dk} for the definition of $g_\ast^{1/2}$. 
$T_f$ is the freeze-out temperature, which is approximately evaluated by
\beq
\frac{ m_\eta }{ T_f } \simeq  \log \Bigl( 0.038 \frac{ M_{\text{pl}} }{ \sqrt{g_\ast} } m_\eta \left< \sigma v \right>_0 \Bigr) 
+ \frac{ 1 }{ 2 } \log \Bigl\{ 
	\log \Bigl( 0.038 \frac{ M_{\text{pl}} }{ \sqrt{g_\ast} } m_\eta \left< \sigma v \right>_0 \Bigr)
\Bigr\}, 
\eeq
where $M_{\text{pl}}$ is the Planck mass, and $\left< \sigma v \right>_0$ is the zeroth term in the expansion of $\left<\sigma v\right>$ by $x^{-1}$~\cite{Kolb:1990vq}. 
The approximate formula in Eq.~(\ref{eq: Relic_density}) is expected to work in about $20~\%$ accuracy~\cite{Belanger:2013ywg}. 

The DM-nucleon scattering is searched for by several experimental groups, and it is severely constrained~\cite{XENON:2018voc, PandaX-4T:2021bab, LZ:2022ufs}. 
In the model, $\eta N \to \eta N$, where $N$ is the nucleon, is generated via a t-channel diagram mediated by the neutral Higgs bosons. 
The spin-independent cross section is approximately given by 
\beq
\sigma_{\mathrm{SI}}\simeq 
\frac{ m_N^2 v^2 }{ 4 \pi (m_\eta + m_N)^2 }
\sum_{a, b = 1}^3
\left( \frac{ g_s^a \lambda_{\eta \eta a}^{} }{ m_{H_a}^2 } \right)
\left( \frac{ g_s^b \lambda_{\eta \eta b}^{} }{ m_{H_b}^2 } \right), 
\eeq
where $m_N$ is the nucleon mass. 
$g_s^a$ ($a=1,2$ and $3$) are coupling constants of scalar couplings between the Higgs bosons and nucleons. 
According to Ref.~\cite{Shifman:1978zn}, they are roughly estimated by 
\beq
\label{eq: Higgs_nucleon_coupling}
g_s^a \simeq
\frac{ 2 }{ 27 } \frac{ m_N^{} }{ v } \bigl( Z_c^a + Z_b^a + Z_t^a \bigr), 
\eeq
where $Z_c^a$, $Z_b^a$ and $Z_t^a$ are defined in Eq.~(\ref{eq: Z_Xfactor}). 
For the SM-like Higgs boson $H_1$, we obtain $g_s^1 \simeq 10^{-3}$.

\section{Baryon asymmetry of the universe}
\label{sec: EWBG}

In this section, we discuss baryogenesis in the model. 
From the Big Bang nucleosynthesis (BBN) and the observation of the light elements of the universe, the ratio of the baryon and photon density, so-called the baryon-to-photon ratio $\eta_B$, is observed as
\beq
\label{eq: EtaB_BBN}
5.8 \times 10^{-10} < \eta_B < 6.5 \times 10^{-10} \quad (\text{BBN}), 
\eeq
with $95~\%$ C.L.~\cite{Fields:2019pfx}.
The cosmic microwave background (CMB) observation also gives a consistent result 
\beq
\label{eq: EtaB_CMB}
6.04 \times 10^{-10} < \eta_B < 6.20 \times 10^{-10} \quad (\text{CMB}),
\eeq
with $95~\%$ C.L.~\cite{Planck:2018vyg}. 
These values mean an unbalance between baryons (matter) and anti-baryons (anti-matter). 
Considering the cosmic inflation, this asymmetry must not have existed from the beginning but must have been created in some era of the early universe. 
Such a scenario is called baryogenesis. 

\subsection{Scenarios of baryogenesis in this model}

For successful baryogenesis, the Sakharov conditions~\cite{Sakharov:1967dj} have to be satisfied: 
(i) the existence of the baryon number violation, (ii) C and CP violation, and (iii) departure from thermal equilibrium. 
In our model, there are roughly two ways to generate the nonzero baryon number satisfying the Sakharov conditions, namely EWBG and leptogenesis. 

In the EWBG~\cite{Kuzmin:1985mm}, the Sakharov conditions are satisfied as follows. 
The baryon number conservation is broken by the sphaleron transition at high temperatures~\cite{Klinkhamer:1984di}. 
The electroweak gauge interaction violates C symmetry. 
The CP-violating source is provided in the Yukawa interaction and/or the Higgs potential. 
Finally, the non-equilibrium state is realized by the strongly first-order electroweak phase transition. 

It is known that EWBG cannot work in the SM because the EWPT in the SM is not strongly first-order but the crossover transition~\cite{Kajantie:1996mn, DOnofrio:2015gop}. In addition, it is known that CP-violation in the SM is too small to generate enough baryon asymmetry even if we assume the strongly first-order EWPT~\cite{Huet:1994jb, ref: GavelaEWBG}. 
On the other hand, in the present model, the Higgs potential is extended at the TeV scale. 
The strongly first-order EWPT can thus occur. 
In addition, new CP-violating sources are provided in the Higgs potential and the Yukawa interaction. 
Therefore, the observed baryon number can be successfully generated via EWBG. 

In leptogenesis~\cite{Fukugita:1986hr}, the non-zero lepton number is generated via the out-of-equilibrium decay of the Majorana fermions~$N^\alpha_R$. 
C and CP-violation are supplied by the phases in the Yukawa matrix~$h_i^\alpha$. 
The generated lepton number is transformed into the baryon number by the sphaleron transition. 
In leptogenesis, super heavy Majorana fermions are necessary~\cite{Davidson:2002qv} if we do not assume the unnatural degeneracy in the mass spectrum of $N^\alpha_R$~\cite{ref: resonant_leptogenesis}. 
However, in the model, such super heavy $N^\alpha_R$ makes neutrino mass too small to explain the experimental data because it is generated at three-loop level. 

Therefore, in this paper, we focus on EWBG, which is more natural as a scenario of baryogenesis in the present model. 
In the model, the strongly first-order EWPT can be realized by non-decoupling quantum effects of the additional scalar bosons like in THDMs~\cite{Turok:1991uc, Cline:1996mga, Kanemura:2004ch, Kanemura:2022ozv}.
EWBG in THDMs has been investigated in Refs.~\cite{Turok:1990zg, Cline:1995dg, Fromme:2006cm, Cline:2011mm, Enomoto:2021dkl, Enomoto:2022rrl}.
It is known that for successful EWBG in THDMs, it is necessary to consider a scenario where a cancellation mechanism works to suppress the eEDM without assuming tiny CP-violating phases~\cite{Kanemura:2020ibp, Bian:2014zka, Fuyuto:2019svr, Cheung:2020ugr}. 
One example of such a scenario is proposed in Ref.~\cite{Kanemura:2020ibp}, 
where the eEDM can be suppressed by the destructive interference between the CP-violating phase in the Yukawa interaction and that in the Higgs potential. 
In addition, in Refs.~\cite{Enomoto:2021dkl, Enomoto:2022rrl}, it has been shown that there are some allowed regions in the THDM for the successful EWBG along this scenario. 
The model in this paper is an extension of the THDMs. 
Therefore, the above cancellation mechanism is also applicable in this model. 

\subsection{Electroweak phase transition}
\label{sec: EWPT}

In this section, we show formulae for the effective potential at finite temperatures of the model, which is necessary to evaluate the EWPT. 
We focus on field configurations maintaining the electromagnetic and $Z_2$ symmetries. 
We thus consider only the classical fields of the Higgs doublets given by
\beq
\label{eq: classical fields}
\Phi_1^\mathrm{cl} = \frac{ 1 }{ \sqrt{2} } 
\begin{pmatrix}
0 \\
\varphi_1 \\
\end{pmatrix}, 
\quad 
\Phi_2^\mathrm{cl} = \frac{ 1 }{ \sqrt{2} }
\begin{pmatrix}
0 \\
\varphi_2 + i \varphi_3 \\
\end{pmatrix}. 
\eeq
The imaginary part of $\Phi_1^\mathrm{cl}$ can be zero by an appropriate gauge fixing. 

The one-loop effective potential $V_\mathrm{eff}$ is given by the sum of the tree-level effective potential $V_0$, one-loop corrections $V_1$ and counterterms. 
$V_0$ is given by substituting the classical fields into the Higgs potential given in Eq.~(\ref{eq:potential_Higgsbasis}). 
It is the same as that in the general THDMs with $\lambda_6 = \mu_{12}^2 = 0$~\cite{Enomoto:2021dkl, Enomoto:2022rrl, Cline:2011mm}. 
$V_1$ is described by the Coleman-Weinberg potential~\cite{Coleman:1973jx};
\beq
\label{eq: CW_potential}
V_1 = \sum_{a}
\frac{ s_a n_a }{ 64 \pi^2 } \tilde{m}^4_a 
\Bigl[
	\log \frac{ \tilde{m}^2_a }{ Q^2 } - \frac{ 3 }{ 2 } 
\Bigr], 
\eeq
where $Q$ is the renormalization scale.
The index $a$ denotes a particle in each loop diagram. 
We employ the Landau gauge, where $V_1$ includes the loop diagrams of the SM fermions, $Z$, $W^\pm$, $\gamma$, the physical scalar bosons and the NG bosons. 
$\tilde{m}_a$ and $n_a$ are the field-dependent mass and the degree of freedom of the particle $a$, respectively.
Formulae for the field-dependent masses are shown in Appendix~\ref{app: field-dependent_mass}.
The statistical factor is denoted by $s_a$ in Eq.~(\ref{eq: CW_potential}), which is one for bosons and minus one for fermions.

To fix counterterms, 
we impose the following renormalization conditions
\beq
\label{eq: RG_condition}
\left( \frac{ \partial V_\mathrm{eff} }{ \partial \varphi_i } \right)_0 = 0, \quad 
\left( \frac{ \partial^2 V_\mathrm{eff} }{ \partial \varphi_i \partial \varphi_j } \right)_0
= (M_h^2)_{ij}, 
\eeq
with $i,j = 1,2,3$ on the 1-loop effective potential $V_\mathrm{eff}$, which is the sum of $V_0$, $V_\mathrm{CW}$ and the counterterms~\cite{Cline:2011mm}. 
The subscript $0$ means that the derivatives are evaluated at the zero temperature vacuum ($\varphi_1^{} = v$ and $\varphi_2^{} = \varphi_3^{} = 0$). 
These renormalization conditions fix the counterterms except for those of $\mu_2^2$, $\lambda_2$, and $\lambda_7$. 
These three counterterms are determined by the $\overline{\mathrm{MS}}$ scheme~\cite{Cline:2011mm}. 

Next, we evaluate the finite temperature correction $V_T$ at one-loop level.
We employ the Parwani scheme for the resummation~\cite{Parwani:1991gq}.  
Then, $V_T$ at temperature $T$ is given by~\cite{Dolan:1973qd} 
\beq
V_T = \sum_a
\frac{ s_a n_aT^4 }{ 2 \pi^2}
\int_0^\infty \mathrm{d}x\, 
\log \biggl\{
	1 - s_a 
	\exp\Bigl(- \frac{\epsilon_a}{ T }\Bigr)
\biggr\}, 
\eeq
where $\epsilon_a = \sqrt{ x^2 + \hat{m}_a^2/T}$ and $\hat{m}_a^2$ is the field-dependent mass of the particle $a$ including the finite temperature correction. 
Formulae for $\hat{m}_a^2$ are shown in Appendix~\ref{app: field-dependent_mass}.

If EWPT is a first-order phase transition, the vacuum bubble is created all over the universe. The bubble profiles of the critical bubble are given by the solutions of the following differential equation;
\beq
\label{eq: bubble_profile}
\frac{ \mathrm{d}^2 \varphi_i }{ \mathrm{d} r^2 } 
+ \frac{ 2 }{ r } \frac{ \mathrm{d} \varphi_i }{ \mathrm{d} r }
= \frac{ \partial (V_\mathrm{eff}+ V_T) }{ \partial \varphi_i }, \quad (i=1,2,3), 
\eeq
with the boundary conditions 
\beq
\varphi_i(\infty) = 0, \quad 
\left. \frac{\mathrm{d}\varphi_i}{\mathrm{d}r} \right|_{r=0} = 0, 
\eeq 
where $r$ is the radial coordinate of the bubble~\cite{Anderson:1991zb}. 

The probability of the bubble nucleation at temperature $T$ is evaluated as~\cite{Anderson:1991zb}
\beq
\Gamma \simeq A(T) \exp\left( - \frac{ S_E }{ T } \right), 
\eeq
where $A(T)\simeq T^4$. 
$S_E$ is the Euclidian action given by 
\beq
S_E = 4 \pi \int_0^\infty \mathrm{d}r\ r^2 
\biggl[
	\sum_i \frac{ 1 }{ 2 } \left( \frac{ \partial \varphi_i }{ \partial r } \right)^2
	+ V_\mathrm{eff} + V_T 
\biggr], 
\eeq
where $\varphi_i$ is the solution of Eq.~(\ref{eq: bubble_profile}). 
The nucleation temperature $T_n$, at which one bubble is expected to exist in each Hubble volume, is roughly evaluated by $S_3/T|_{T=T_n} \simeq 140$~\cite{Anderson:1991zb}. 

\subsection{Electroweak baryogenesis}
\label{sec: EWBG2}

In this section, we discuss EWBG in the model. 
We consider the case that the baryon asymmetry is generated by the charge transport mechanism~\cite{Nelson:1991ab}. 
We shortly review it in the following. 

In the first order EWPT, vacuum bubbles are created all over the universe. 
Bubbles with a large enough radius expand, and the broken phase finally fills the entire universe.\footnote{If the nucleation rate is too small, 
it can happen that there is no solution of $S_3/T|_{T=T_n} \simeq 140$. 
In such a case, EWPT has not been completed until the present~\cite{Apreda:2001us}.} 
This bubble expansion makes the universe far from thermal equilibrium. 
The bubble wall interacts with plasmas in the thermal bath during the expansion. 
If this interaction includes enough C and CP-violation, 
the charge accumulations with opposite signs are generated in the front and the back of the wall, respectively by the reflection and the penetration of the plasmas~\cite{Nelson:1991ab}. 
The accumulated charge of the left-handed fermions is converted into the baryon asymmetry by the sphaleron transition~\cite{Klinkhamer:1984di}. 
In the strongly first-order EWPT, 
the sphaleron transition can occur outside of the bubble (the symmetric phase), on the other hand, it rapidly decouples inside the bubble (the broken phase) due to nonzero VEV. 
Then, the generated baryon asymmetry in the front of the bubble wall freezes out in the broken phase and remains in the present universe. 
As explained above, in order to calculate the baryon number density,
the interaction between the bubble wall and thermal plasmas is so important in the charge transport mechanism. 
Thus, we first discuss which particles are in thermal plasmas and should be included in the evaluation of baryon asymmetry. 
In the charge transport mechanism, a heavier SM fermion gives a larger contribution because it more strongly interacts with the Higgs doublets, i.e. the bubble wall. 
Thus, the top quark gives the main contribution~\cite{Fromme:2006wx}.   
Although the other quarks are light, they interact with the top quark via QCD processes. 
Therefore, they should also be included in the evaluation. 
On the other hand, the leptons can be neglected~\cite{Cline:2000nw, Fromme:2006wx, Fromme:2006cm}.
 
We assume that $m_{N^\alpha}^{}$'s are a few TeV, and the masses of the other new particles in the model are around the electroweak scale. 
This is the case of the benchmark scenario shown in Sec.~\ref{sec: Benchmark}. 
$N^\alpha_R$ are thus absent from the thermal bath at the EWPT. 
On the other hand, the scalar bosons can be included in the thermal bath. 
The Higgs doublets interact with the top quark via the Yukawa interaction. 
It affects the charge accumulation. 
Therefore, we include both the Higgs doublets in the evaluation. 
$S^\pm$ interact with only the leptons at tree level. 
In addition, the interaction between $S^\pm$ and the leptons receives a suppression by $m_{N^\alpha}^{}$. 
Thus, the effect of $S^\pm$ are expected to be negligibly small. 
$\eta$ does not have interaction with the SM fermions at tree level. 
$\eta$ is also expected to give no considerable contribution to baryon asymmetry.  
Consequently, in the evaluation of the charge accumulations, 
we consider the quarks and the Higgs doublets in the thermal bath. 
It is the same situation of EWBG in the THDMs~\cite{Turok:1990zg, Cline:1995dg, Fromme:2006cm, Cline:2011mm, Enomoto:2021dkl, Enomoto:2022rrl}. 

In evaluating the charge accumulation, we employ the WKB method~\cite{Joyce:1994fu, Cline:2000nw, Fromme:2006wx, Cline:2020jre}, where the typical momentum of the plasmas is assumed to be sufficiently larger than $1/L_w$, where $L_w$ is the bubble wall width. Thus, $L_w$ has to be larger enough than $1/T$ for the WKB method to be valid, where $T$ is the typical temperature of EWPT. 
In addition, we assume that the bubble wall velocity $v_w$ is so small that the relativistic effects can be neglected~\cite{Fromme:2006cm, Cline:2011mm, Enomoto:2021dkl}. 
We use formulae in Ref.~\cite{Enomoto:2021dkl}, where the same situation is considered. 
See Ref.~\cite{Enomoto:2021dkl} for explicit formulae. 

Once the charge accumulation is found, 
$n_B/s$ is given by integrating the charge accumulation of the left-handed fermions in the front of the bubble wall considering the sphaleron transition rate~\cite{Cline:2011mm}, where $n_B$ and $s$ are the baryon number and entropy density, respectively. 
The baryon-to-photon ratio $\eta_B$ is then given by $\eta_B^{} \simeq 7.04 \times (n_B/s)$~\cite{PDG2022}. 

\section{A benchmark scenario}
\label{sec: Benchmark}

In this section, we introduce a benchmark scenario of the model, where neutrino mass, dark matter, and the BAU can be explained simultaneously.
The input parameters are given as follows. 
\begin{itemize}
\item {\bf Masses of the new particles}
\bal
& m_{H^\pm}^{} = m_{H_3}^{} = 250~\mathrm{GeV}, \quad 
m_{H_2} = 420~\mathrm{GeV}, \quad 
m_S = 400~\mathrm{GeV}, \quad m_\eta = 63~\mathrm{GeV}, \nonumber \\
& (m_{N^1}, m_{N^2}, m_{N^3}) = (3000, 3500, 4000)~\mathrm{GeV}. \nonumber 
\eal
\item {\bf Parameters in the Higgs potential}
\bal
&\mu_2^2 = (50~\mathrm{GeV})^2, \quad \mu_{12}^2 = 0, \quad 
\mu_S^2 = (320~\mathrm{GeV})^2, \quad 
\lambda_2 = 0.1, \quad \lambda_6 = 0, \nonumber \\[3pt]
& |\lambda_7| = 0.817, \quad \theta_7  = -0.730, \quad 
 |\rho_{12}| = 0.1, \quad \theta_\rho  = -2.94, \quad 
\rho_2 = 0.1, \nonumber \\[3pt]
& \sigma_{1} = |\sigma_{12}| = 1.1 \times 10^{-3}, \quad \theta_\sigma = 0, \quad 
\sigma_2 = 0.1, \quad  \kappa = 2, \quad 
\lambda_S^{} = 1, \quad \lambda_\eta = 1, \quad \xi = 1. \nonumber 
\eal
\item {\bf Parameters in the Yukawa interaction}
\bal
& \frac{ m_u }{ v } |\zeta_u | = 2.15 \times 10^{-6}, \quad 
	\frac{ m_c }{ v } |\zeta_u | = 1.26 \times 10^{-3}, \quad 
	\frac{ m_t }{ v } |\zeta_u| = 0.172, \nonumber \\[5pt]
& \frac{ m_d }{ v } |\zeta_d| = 4.65 \times 10^{-6}, \quad 
	\frac{ m_s }{ v } |\zeta_d| = 9.31 \times 10^{-5}, \quad 
	\frac{ m_b }{ v } |\zeta_d| = 4.16 \times 10^{-3}, \nonumber \\[5pt]
& \frac{ m_e }{ v } |\zeta_e| = 2.5 \times 10^{-4}, \quad 
	\frac{ m_\mu }{ v } |\zeta_\mu| = 2.5 \times 10^{-4}, \quad 
	\frac{ m_\tau }{ v } |\zeta_\tau| = 2.5 \times 10^{-3}, \nonumber \\[5pt]
& \theta_u = \theta_d = 0.245, \quad 
	\theta_e = \theta_\mu = \theta_\tau = -2.946,\nonumber  
\eal
\bal
& 
\begin{pmatrix}
h^1_1 & h^1_2 & h^1_3 \\[3pt]
h^2_1 & h^2_2 & h^2_3 \\[3pt]
h^3_1 & h^3_2 & h^3_3 \\
\end{pmatrix}
=
\begin{pmatrix}
1.00 \, e^{-0.314i} & 0.196  \, e^{0.302i} & 1.04 \, e^{-2.39i} \\[3pt]
1.08 \, e^{-1.88i} & 0.205 \, e^{-1.80i} & 1.05 \, e^{2.33i} \\[3pt]
0.449 \, e^{2.74i} & 1.31 \, e^{-0.0331i} & 0.100 \, e^{0.628i} \\
\end{pmatrix}. \nonumber 
\eal
\end{itemize}
By using these input values, 
the remaining scalar couplings are determined as 
\beq
\mu_\eta^2 \simeq (62.7~\mathrm{GeV})^2, \quad 
\lambda_3 \simeq 1.98, \quad \lambda_4 = \lambda_5 \simeq 1.88, \quad 
\rho_1\simeq 1.90. 
\eeq

We can see that all the theoretical constraints in Sec.~\ref{sec: theoretical_constraints} can be avoided in the benchmark scenario. 
According to Ref.~\cite{Dorsch:2016nrg}, the Landau pole is around $5$--$10~\mathrm{TeV}$ in the benchmark scenario. 
At the scale above the Landau pole, 
a new UV theory is expected to appear including the present model as an effective low-energy theory. 
An example of such a UV theory has been investigated in Ref.~\cite{ref: UVextension}, 
where the Higgs bosons are composite states of heavy fermions confined by the non-perturbative effect of an additional SU(2) gauge symmetry. 
More details on the Landau pole and a UV extension of the present model will be studied elsewhere~\cite{ref: futureAKS}. 

We also have checked that the benchmark scenario avoids the constraints from the high-energy colliders and the flavor experiments explained in Sec.~\ref{sec: experimental_constraints}. 
The branching ratios of the lepton flavor violating decays and their experimental upper bounds are shown in Table~\ref{table: LFV_benchmark}. 
We can see that the experimental constraint can be avoided in all the lepton flavor-violating decays. 

The CP-violating phases $\theta_e$ and $\theta_\rho$ have the same value in the benchmark scenario.
The Barr-Zee type diagram including $S^\pm$ loop gives no contribution to the eEDM because it is proportional to $\sin (\theta_e - \theta_\rho)$. 
Then, the eEDM in the benchmark scenario can be evaluated by using the formula in Ref.~\cite{Kanemura:2020ibp} at one-loop level. 
The result is $d_e = 0.22 \times 10^{-30}~\mathrm{e\, cm}$ which is less than the latest upper bound $|d_e| < 4.1 \times 10^{-30}~\mathrm{e\, cm}$~\cite{Roussy:2022cmp}. 
Consequently, the benchmark scenario can avoid all the current experimental constraints discussed in Sec.~\ref{sec: experimental_constraints}. 

With the above input values, 
the model can reproduce the neutrino mass parameters in the normal ordering case derived in the global fit by the Particle Data Group~\cite{PDG2022}; 
\beq
\begin{array}{l}
 \Delta m_{21}^2 = 7.53 \times 10^{-5}~\mathrm{eV^2}, \quad 
\Delta m_{32}^2 = 2.453 \times 10^{-3}~\mathrm{eV^2}, \\
\sin^2 \theta_{12} = 0.307,  \quad \sin^2 \theta_{23} = 0.546, \quad 
\sin^2 \theta_{13} = 2.20 \times 10^{-2}, \quad \delta = 1.36 \pi, 
\end{array}
\eeq
where the notation follows Ref.~\cite{PDG2022}.
In addition, the Majorana phases in the PMNS matrix are predicted to be $\alpha_1 = 0$, $\alpha_2 = -\pi /2$.  
The lightest neutrino mass is $m_{\nu^1} \simeq 0.006~\mathrm{eV}$. 
Thus, the sum of the neutrino mass is about $0.067~\mathrm{eV}$, 
which is lower than the current upper bound from the CMB observation~\cite{Planck:2018vyg}. 

The effective Majorana neutrino mass $\left< m_{\beta \beta} \right> = | (M_\nu)_{ee} |$ induces the neutrinoless double beta decay ($0\nu \beta \beta$), which violates the lepton number conservation~\cite{Umehara:2008ru, NEMO-3:2009fxe, NEMO-3:2016qxo, Barabash:2018yjq, GERDA:2020xhi, CUPID:2022puj, Augier:2022znx, CUORE:2022piu, CUORE:2021mvw, KamLAND-Zen:2022tow}. 
Ignoring the difference in isotopes, 
the strongest bound is given by the KamLAND-Zen experiment; $\left< m_{\beta \beta}^{} \right> < 36\text{--}156~\mathrm{meV}$~\cite{KamLAND-Zen:2022tow}. 
In the benchmark scenario, $ \left< m_{\beta \beta} \right>$ is about $1~\mathrm{meV}$, and it avoids the current constraint. 
In the present model, $0\nu\beta \beta$ is also caused by a one-loop induced dimension-five operator $e_R^- e_R^- H^+ H^+$ described by red subdiagrams of the neutrino mass diagrams in Fig.~\ref{fig: Neutrino_mass_diagram}. In the benchmark scenario, the coefficient of this operator is roughly evaluated by $C/\Lambda \simeq 2 \times 10^3~\mathrm{TeV}^{-1}$, where $C \simeq (16\pi^2)^{-1}$ and $\Lambda \simeq m_{N^1}^{} \simeq 3~\mathrm{TeV}$. Although it seems to give a larger contribution to $0 \nu \beta \beta$, we found that it gives only two to three orders of magnitude smaller contributions to the amplitude of $0\nu\beta\beta$ than the active-neutrino-mediated contribution because of the suppression by $m_{H^\pm}^{} > m_W^{}$ and small couplings between $H^\pm$ and nucleons roughly evaluated by $10^{-3} \times |\zeta_q^{}|$ (See Eq.~(\ref{eq: Higgs_nucleon_coupling})). Therefore, the benchmark scenario avoids the current constraint from observations of $0\nu\beta \beta$.

\begin{table}[t]
\begin{center}
\begin{tabular}{|c|c|c|}\hline
\ Process\  & \ Branching ratio\  & \ Upper bound\   \\ \hline
$\mu \to e \gamma$ & $1.4 \times 10^{-14}$ & $4.2 \times 10^{-13}$~\cite{MEG:2016leq}  \\ 
$\tau \to e \gamma$ & $5.3 \times 10^{-10}$ & $3.3 \times 10^{-8}$~\cite{BaBar:2009hkt}  \\ 
$\tau \to \mu \gamma$ & $1.1 \times 10^{-11}$ & $4.4 \times 10^{-8}$~\cite{BaBar:2009hkt}  \\ \hline
\end{tabular}
\hspace{0.05\textwidth}
\begin{tabular}{|c|c|c|}\hline
\ Process\  & \ Branching ratio\  & \ Upper bound\   \\ \hline
$\mu \to 3 e$ & $1.0 \times 10^{-13}$ & $1.0\times 10^{-12}$~\cite{SINDRUM:1987nra}   \\ 
$\tau \to 3 e$ & $6.2 \times 10^{-10}$ & $2.7 \times 10^{-8}$~\cite{Hayasaka:2010np}   \\ 
$\tau \to 3 \mu$ & $2.4 \times 10^{-11}$ & $2.1 \times 10^{-8}$~\cite{Hayasaka:2010np}  \\ 
$\tau \to e \mu \overline{e}$ & $5.1 \times 10^{-12}$ & $1.8 \times 10^{-8}$~\cite{Hayasaka:2010np}  \\ 
$\tau \to \mu \mu \overline{e}$ & $1.1\times 10^{-12}$ & $1.7 \times 10^{-8}$~\cite{Hayasaka:2010np}  \\ 
$\tau \to e e \overline{\mu}$ & $4.5 \times 10^{-13}$ & $1.5\times 10^{-8}$~\cite{Hayasaka:2010np}  \\ 
$\tau \to e \mu \overline{\mu}$ & $9.6 \times 10^{-11}$ & $2.7 \times 10^{-8}$~\cite{Hayasaka:2010np}  \\ \hline
\end{tabular}
\end{center}
\caption{The branching ratio for the lepton flavor violating decays in the benchmark scenario and the current experimental bounds at $95~\%$ C.L.}
\label{table: LFV_benchmark}
\end{table}

We have also checked that the benchmark scenario can reproduce the observed DM relic abundance $\Omega_{\text{DM}} h^2 \simeq 0.12$. 
For the value of $g_\ast^{1/2}$ in Eq.~(\ref{eq: Relic_density}), we have used the result in Ref.~\cite{Drees:2015exa}.
The spin-independent cross section in the benchmark scenario is $\sigma_\mathrm{SI} = 2.3 \times 10^{-48}~\mathrm{cm^2}$. 
It is lower than the current upper bound~\cite{XENON:2018voc, PandaX-4T:2021bab, LZ:2022ufs}. 

\begin{figure}[b]
\begin{center}
\includegraphics[width=0.6\textwidth]{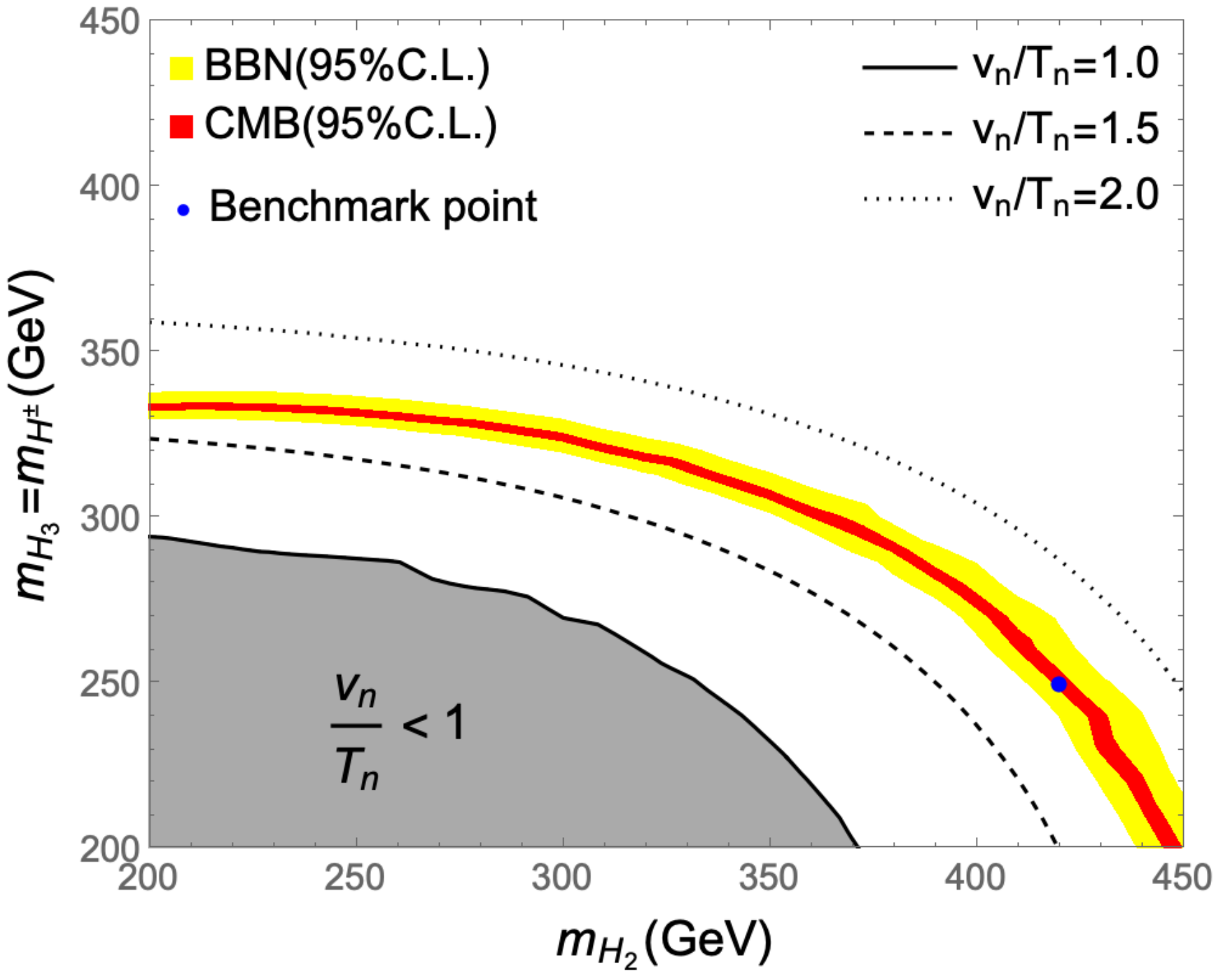}
\caption{The baryon asymmetry in the model. 
Parameters but $m_{H_2}^{}$, $m_{H_3}^{}$($= m_{H^\pm}^{}$) are fixed to the benchmark values. 
In the red (yellow) region, the observed baryon asymmetry from CMB (BBN) can be explained within $95~\%$ C.L. 
Contours for $v_n/T_n = 1$, $1.5$ and $2.0$ are shown with the black solid, dashed and dotted lines, respectively. 
In the gray region, $v_n/T_n$ is less than one, and the EWPT is not the strongly first-order one although it is still the first-order one. 
The blue point corresponds to the benchmark scenario.}
\label{fig: BAUScanPlot}
\end{center}
\end{figure}

Finally, we evaluate the baryon asymmetry. 
In Fig.~\ref{fig: BAUScanPlot}, we show the result of the numerical evaluation of baryon asymmetry for various masses of the additional Higgs bosons. 
In the evaluation, we set $v_w$ to be $0.1$~\cite{Fromme:2006cm, Enomoto:2021dkl, Cline:2011mm}. 
$m_{H_3}^{}$ and $m_{H^\pm}^{}$ are degenerated.  
The model parameters but $m_{H_2}^{}$, $m_{H_3}^{}$ and $m_{H^\pm}^{}$ are fixed to the values in the benchmark scenario. 
We have used \texttt{CosmoTransitions}~\cite{Wainwright:2011kj} for the evaluation of the EWPT. 
The renormalization scale $Q$ is set to be $m_Z^{}$. 
We employ the nucleation temperature $T_n$ as the typical temperature of the EWPT. 
The black solid, dashed and dotted lines are contours for $v_n/T_n = 1$, $1.5$ and $2.0$, respectively, where $v_n$ is the VEV at the nucleation temperature. 
In the gray region, although the EWPT occurs in the first order, 
$v_n/T_n$ is less than one, i.e., the EWPT is not the strongly first-order one.
$v_n/T_n$ is larger in heavier mass regions. 
It is because that $\mu_2^2$ is fixed in the evaluation. 
Then, the non-decoupling effect of the additional Higgs bosons is enhanced by heavier masses.  
The region for successful EWBG is shown in the red and yellow regions. 
In the red (yellow) region, the observed baryon-to-photon ratio from CMB (BBN) can be explained within $95~\%$ C.L. 
In these regions, the WKB method gives a valid approximation because the bubble wall width is estimated as $5/T_n$. 
In heavier mass regions, the baryon asymmetry is thus overproduced. In such regions, the bubble wall width is smaller, and the WKB method would not be valid. 
The blue point corresponds the benchmark scenario $(m_{H_2}^{}, m_{H_3}^{}, m_{H^\pm}) = (420, 250, 250)~\mathrm{GeV}$. 

\begin{figure}[t]
\begin{center}
\includegraphics[width=0.7\textwidth]{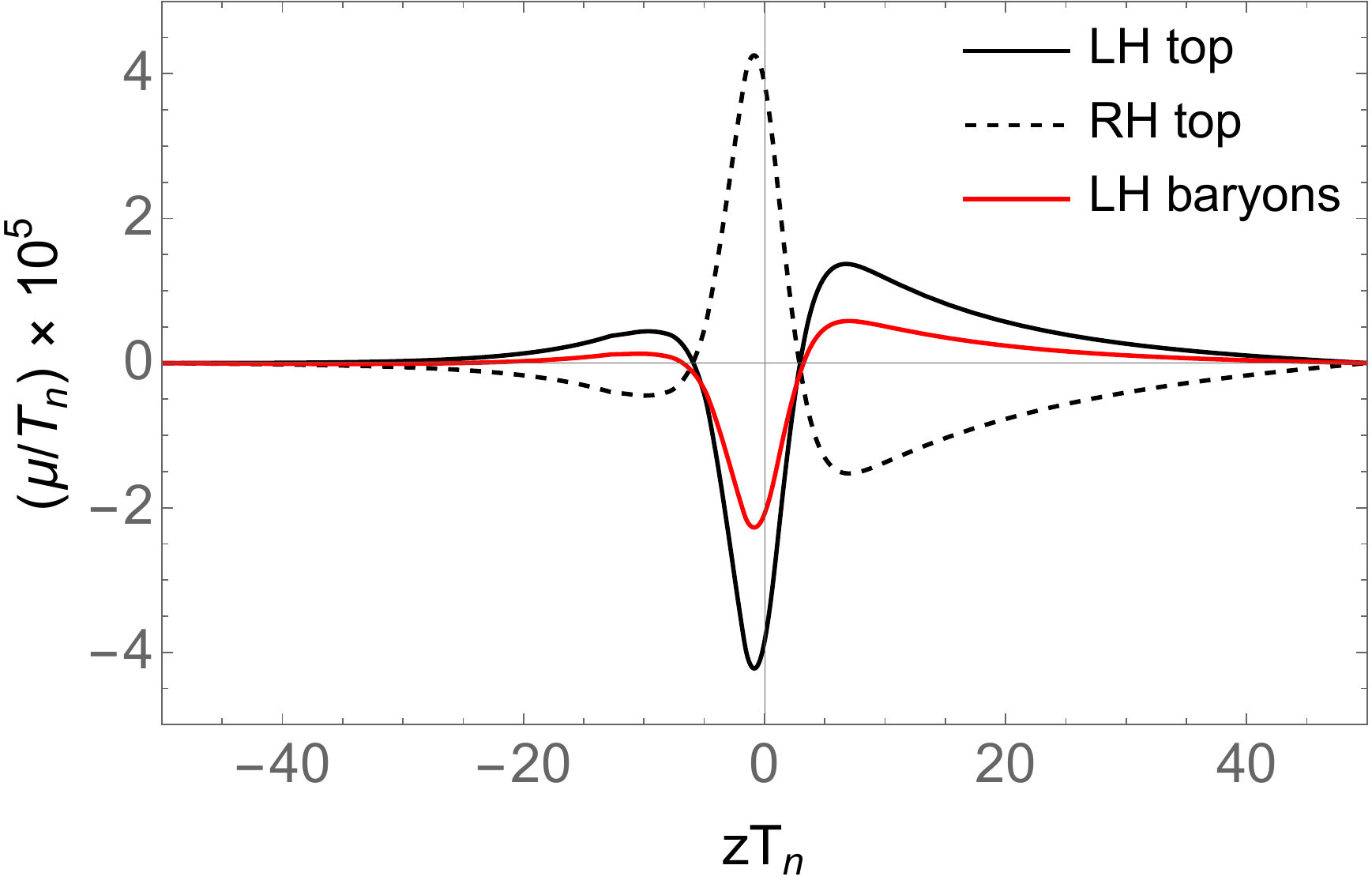}
\caption{The distributions of the chemical potentials of the left-handed (LH) top quark, right-handed (RH) top quark and the LH baryons are shown in black-solid, black-dashed and red-solid lines, respectively.}
\label{fig: Chemical_potential}
\end{center}
\end{figure}
In the benchmark scenario, 
the nucleation temperature is $T_n \simeq 100~\mathrm{GeV}$, and $v_n / T_n \simeq 1.7$. 
In Fig.~\ref{fig: Chemical_potential}, we show the distributions of the chemical potentials normalized by $T_n$ in the benchmark scenario. 
The horizontal axis of the figure is the radial coordinate $z$ of the vacuum bubble normalized by the nucleation temperature $T_n$ in the wall frame, where the bubble wall is stationary at $z=0$. 
The positive (negative) direction of $z$ is directed to the symmetric (broken) phase.  
The main source of the charge asymmetry is given by the charge transport of top quarks. 
The chemical potentials of the left-handed (LH) and right-handed (RH) top quarks are shown in black-solid and black-dashed lines, respectively. 
The source of baryon asymmetry is given by the chemical potential of the LH baryons $\mu_{B_L^{}}^{}$~\cite{Cline:2000nw}; 
\begin{equation}
\mu_{B_L^{}}^{} = \frac{ 1 }{ 2 } \sum_q \mu_{q_L}^{}. 
\end{equation}
$\mu_{B_L^{}}^{}$ in the symmetric phase ($z>0$) is transformed into 
non-zero baryon number by the sphaleron transition. 
$\mu_{B_L^{}}^{}$ in the benchmark scenario is shown in the red-solid line in Fig.~\ref{fig: Chemical_potential}. 
The baryon-to-photon ratio in the benchmark scenario is evaluated as $\eta_B = 6.17 \times 10^{-10}$.
It is consistent with observed values from CMB and BBN. 

Consequently, the benchmark scenario can explain the neutrino mass, dark matter and the BAU simultaneously with satisfying the theoretical and experimental constraints. 
In the following, we shortly discuss the prediction of the benchmark scenario in several future experiments. 

The charged Higgs boson in the benchmark scenario is dominantly produced at hadron colliders via $gg \to H^\pm tb$ and decays into $tb$ at almost $100~\%$. 
For $m_{H^\pm} = 250~\mathrm{GeV}$, 
the future HL-LHC with the integrated luminosity $L = 3000~\mathrm{fb^{-1}}$ is expected to verify the signal up to $|\zeta_u| \lesssim 0.2$~\cite{Aiko:2020ksl, Arbey:2017gmh, Arhrib:2018ewj}.
In the benchmark scenario, $|\zeta_u| \simeq 0.25$, and $H^\pm$ would be detected in the HL-LHC. 

$H_2$ and $H_3$ in the benchmark scenario is predominantly produced via the gluon fusion $gg \to H_{2,3}$ at hadron colliders. 
$H_2$ decays into $t\bar{t}$ at almost $100~\%$. 
On the other hand, $H_3 \to t \bar{t}$ is kinematically prohibited. 
It mainly decays into $gg$, $b\bar{b}$ and $\tau \bar{\tau}$. 
These signals would also be tested at the future HL-LHC. 

A pair of $S^\pm$ is expected to be produced at future $e^+e^-$ and hadron colliders. 
In the benchmark scenario, $S^\pm$ decays into $H^\pm \eta$ at almost $100~\%$, and $H^\pm$ decays into $tb$. 
Thus, the signal from $S^+S^- \to t\bar{t} b\bar{b} \cancel{E}$ is expected. 
The final states depend on the decay of $W^\pm$; $2b 2\bar{b} 4j \cancel{E}$, $2b 2\bar{b} 2j \ell^\pm \cancel{E}$ or $2b 2\bar{b} \ell^- \ell^+ \cancel{E}$. 
This signal would be effective to search for $S^\pm$ at future high-energy colliders. 

In the MEG-II experiment, the expected upper limit on $\mathrm{Br}(\mu \to e \gamma)$ is $6\times 10^{-14}$~\cite{MEGII:2018kmf}. 
Although this is a quite precise measurement, 
the prediction in the benchmark scenario is smaller. 
$\tau \to e \gamma$ and $\tau \to \mu \gamma$ are expected to be tested at the Belle-II experiment. 
The upper limits for them are expected to be improved to $3 \times 10^{-9}$ and $1 \times 10^{-9}$, respectively with the integrated luminosity $L = 50~\mathrm{ab^{-1}}$~\cite{Belle-II:2018jsg}. 
However, the predictions in the benchmark scenario for them are also smaller than these expected upper limits. 
The Mu3e experiment will search for $\mu \to 3e$. 
At phase I and phase II, the expected upper bound on the branching ratio is $2\times 10^{-15}$ and $1 \times 10^{-16}$, respectively. 
The benchmark scenario would be tested at phase I. 
The other three body decays will be searched at the Belle-II experiment. 
The expected upper bounds on the branching ratios are a few $10^{-10}$ with $L = 50~\mathrm{ab^{-1}}$. 
The benchmark scenario would be tested by the observation of $\tau \to 3 e$. 

In the benchmark scenario, the eEDM is evaluated as $|d_e| = 0.22 \times 10^{-30}~\mathrm{e\, cm}$.
The ACME experiment expects that the future improvement can reduce the upper bound to $10^{-30}~\mathrm{e\, cm}$~\cite{ACME:2018yjb}. 
Thus, it might be difficult to detect the eEDM in the benchmark scenario 
at the improved ACME experiment. 
However, more future experiments would be possible to reach the value in the benchmark scenario.
The nEDM measurement is also expected to improve the upper bound by an order of magnitude~\cite{Martin:2020lbx}. 
However, the nEDM in the benchmark scenario is a few orders of magnitude smaller than this expected upper limit. 

The CP-violation in the model would also be possible to be tested at future collider experiments. 
In Ref.~\cite{Kanemura:2021atq}, the collider test of $\theta_\tau$ is investigated in the THDM focusing on the azimuthal angle distributions of a pair of $\tau^+ \tau^-$ as a decay product of the additional neutral Higgs bosons at high-energy $e^+e^-$ colliders. 
This would be useful to test $\theta_\tau$ and $H_{2,3}$ in the present model. 

The KamLAND-Zen experiment has a future plan to improve the $0\nu\beta \beta$ search, KamLAND2-Zen~\cite{Ichimura:2022kvl}. 
It is planned to explore $\left< m_{\beta \beta} \right>$ down to $20~\mathrm{meV}$ in five years and will cover the whole region in the case of the inverted neutrino mass ordering. However, the prediction in the benchmark scenario, which is the normal ordering case, is a smaller value $\left< m_{\beta \beta} \right> \simeq 1~\mathrm{meV}$. 
One more order of magnitude update would be necessary to test it. 

The one-loop induced lepton-number-violating (LNV) operator $e_R^- e_R^- H^+ H^+$ gives only a negligibly small contribution in $0\nu\beta \beta$ as explained above. Thus, it would also be difficult to test it at future $0\nu\beta\beta$ searches. 
However, this kind of operator $\ell^- \ell^{\prime ^-} H^+ H^+$ ($\ell$, $\ell^\prime = e, \mu, \tau$) induces other LNV processes, for example, LNV signals at high-energy colliders. 
Such an effect can be sizable enough to be detected in future experiments because it is not suppressed by the small couplings between $H^\pm$ and nucleons.
In general, such LNV effective operators, which are not directly cause tiny Majorana neutrino mass at tree level but generate it at loop level, 
can induce larger LNV effects than those via the Majorana neutrino mass because of fewer loop suppressions. Hence, they can lead a richer phenomenology at searches for lepton number violation, for example, $0\nu\beta\beta$, $\mu^-$-$e^+$ conversion, LNV rare meson decays and LNV signals at colliders. 
This gives an important prediction in radiative seesaw models and has been investigated in various models~\cite{Zee:1980ai, Cheng:1980qt, Zee:1985id, Babu:1988ki, Krauss:2002px, Ma:2006km, ref: AKS, Gustafsson:2012vj, Aoki:2010tf, Gustafsson:2014vpa, Gustafsson:2020bou, Babu:2022ycv, Cai:2017mow} and by using LNV effective operators~\cite{Babu:2001ex, Angel:2012ug, delAguila:2012nu, deGouvea:2007qla, Cepedello:2017lyo, Herrero-Garcia:2019czj, Fuks:2020zbm, Deppisch:2020oyx, DeGouvea:2019wnq, Aoki:2020til, Berryman:2016slh}. 
Thus, studies on LNV processes via the $\ell^- \ell^{\prime -} H^+ H^+$ operator are also expected to provide an effective way to verify the present model in future experiments and will be investigated elsewhere~\cite {ref: futureAKS}.

The LZ experiment with the full data set is expected to be possible to test $\sigma_\mathrm{SI}$ up to $10^{-48}~\mathrm{cm^2}$~\cite{Kudryavtsev:2019bui}. 
The prediction in the benchmark scenario is $\sigma_\mathrm{SI} = 2.3 \times 10^{-48}~\mathrm{cm^2}$. 
It would be possible to test the DM in the benchmark scenario at the improved LZ experiment. 

In the benchmark scenario, the strongly first-order EWPT is expected to occur by the non-decoupling effect of the additional scalar bosons. 
It has been investigated in several previous works how such a non-decoupling effect is experimentally tested. 
First, it is known that the Higgs triple coupling largely deviates from the SM prediction~\cite{Kanemura:2002vm, Kanemura:2004mg, Kanemura:2004ch}. 
This deviation is expected to be tested at future high-energy colliders. 
At the HL-LHC, the measurement with $50~\%$ accuracies is expected~\cite{Cepeda:2019klc}. 
Furthermore, more precise measurements with $27~\%$ and $10~\%$ accuracies are expected at the future upgraded ILC with the beam energy $500~\mathrm{GeV}$ and $1~\mathrm{TeV}$, respectively~\cite{Bambade:2019fyw}. 

In the present model, this deviation at one-loop level is evaluated by~\cite{Kanemura:2002vm, Kanemura:2004mg} 
\bal
\Delta R = \frac{ 1 }{ 12 \pi^2 v^2 m_{H_1}^2 }
\biggl\{&
	 \frac{ 2 (m_{H^\pm}^2 - \mu_2^2 )^3 }{ m_{H^\pm}^2 }
	+ \frac{ (m_{H_2}^2 - \mu_2^2 )^3 }{ m_{H_2}^2 }
	\nonumber \\[7pt]
	& + \frac{ (m_{H_3}^2 - \mu_2^2 )^3 }{ m_{H_3}^2 }
	+ \frac{ 2 (m_{S}^2 - \mu_S^2 )^3 }{ m_{S}^2 }
	+ \frac{ (m_{\eta}^2 - \mu_\eta^2 )^3 }{ m_{\eta}^2 }
\biggr\}. 
\eal
In the benchmark scenario, $\Delta R$ is predicted to be $38~\%$. 
This would thus be verified at the HL-LHC and the upgraded ILC. 
In evaluating the non-decoupling effect in $\Delta R$, 
two-loop corrections are also important. Two-loop corrections in other extended Higgs sectors are studied in Refs.~\cite{Braathen:2019zoh, Braathen:2020vwo}. 

The non-decoupling effect of the charged scalar bosons can also largely shift the prediction for the diphoton decay of the Higgs boson $H_1 \to \gamma \gamma$~\cite{Ellis:1975ap, Shifman:1979eb}. 
The product of the total production cross section of the Higgs boson and the branching ratio for the diphoton decay is currently observed with the value $\sigma \times \mathrm{Br} = 127\pm 10~\mathrm{fb}$ by using the full Run 2 data~\cite{ATLAS:2020pvn}. 
In the present model, the production cross section of the Higgs boson is almost the same with that in the SM because we assume the alignment structures at tree level explained in Sec.~\ref{sec: Model}. 
The branching ratio for $H_1 \to \gamma \gamma$ receives contributions from 1-loop diagrams of $H^\pm$ and $S^\pm$. 
In the benchmark scenario, it is expected that $\sigma \times \mathrm{Br} = 100 \pm 4~\mathrm{fb}$, where we use the value $116 \pm 5~\mathrm{fb}$ as the total production cross section in the SM~\cite{ATLAS:2020pvn}. 
There is a small overlap in the $2\sigma$ regions of the observed $\sigma \times \mathrm{Br}$ and that in the benchmark scenario. 
It would be tested at near future upgrades at LHC~\cite{Chen:2017khz}. 
If the benchmark scenario will be excluded by future $H_1 \to \gamma \gamma$ measurements, 
the strongly first EWPT would still be possible to occur in the present model by considering the larger non-decoupling effect of the additional neutral scalar bosons $H_2$, $H_3$ and $\eta$. Thus, even in such a case, it is expected that there are some parameter regions of the model where neutrino mass, dark matter and the BAU can be explained simultaneously. 

The strongly first-order EWPT can also be tested by future gravitational wave (GW) observations in space~\cite{Grojean:2006bp, Dorsch:2016nrg, Kakizaki:2015wua, Hashino:2016rvx, Hashino:2018wee, Kanemura:2022txx, Enomoto:2022rrl}. 
During the phase transition, 
GWs are expected to be produced from three sources: the collision of bubble walls~\cite{ref: GW_bubble_collision}, the sound waves in the plasma after the bubble collisions~\cite{Hindmarsh:2013xza} and the magnetohydrodynamic turbulence forming after the bubble collisions~\cite{Caprini:2006jb}. 
Some space-based observatories are currently planned to aim to detect such GWs: LISA~\cite{LISA:2017pwj}, DECIGO~\cite{Seto:2001qf} and BBO~\cite{Corbin:2005ny}. 
In Refs.~\cite{Dorsch:2016nrg, Enomoto:2022rrl}, GWs from the strongly first-order EWPT in THDMs with the CP-violation have been evaluated, and detectable GWs spectra are expected in some parameter regions. 
Since the present model is an extension of THDMs, 
it would also be tested at the future space-based observatory. 
This will be investigated in more detail elsewhere~\cite{ref: futureAKS}. 

Finally, it has recently been studied to test the first-order EWPT in observations of the primordial black holes (PBHs)~\cite{ref: Primordial BHs}. 
According to these references, sizable non-decoupling effects can make the remnant of the PBHs large enough to be probed by the current and future PBH observations. 
This way might be applicable in some parameter regions of the present model. 

Although we show only one benchmark point in this paper, 
it is expected that there are various other parameter points where neutrino mass, dark matter and BAU can be explained under the theoretical and experimental constraints. 
However, an exhaustive examination of the entire parameter space of the model is beyond the scope of this work. It will be performed elsewhere~\cite{ref: futureAKS}.

\section{Discussions and conclusions}
\label{sec: conclusions}

In this paper, we show only one benchmark scenario. 
However, there would be several other possible scenarios. 
For the neutrino mass, the scenario for the inverted ordering case can be considered. 
It would be given by changing the value of $h^\alpha_i$. 
A scenario with Majorana dark matter $N^1_R$ would also be conceivable. 
In that case, $N^1_R$ is so-called the leptophilic dark matter~\cite{Krauss:2002px}. 
A pair of them annihilates into a pair of charged leptons. 
In order to avoid strong LFV limits, the heavy Majorana fermions are favored to be at the TeV scale. 
Thus, in the case that $N^1_R$ is the dark matter particle, all the $Z_2$-odd particles are heavy. 
In such a case, it would be difficult to reproduce the neutrino oscillation data because the three-loop neutrino masses are too small. 


In this paper, we have ignored the effect of the lepton number violation via the Majorana masses of $N_R^\alpha$'s during baryogenesis. 
Considering strongly degenerated Majorana fermions, 
it would be possible to generate the observed baryon asymmetry by the mechanism of resonant leptogenesis. 
In the case that their mass spectra are separated enough, like in the benchmark scenario, the generated lepton number density is expected to be much less than the amount needed to explain the observed baryon asymmetry.
Even in such a case, it would slightly affect the conversion rate of the charge accumulation into baryon asymmetry via the sphaleron transition~\cite{Cline:2000nw}. 
This possibility will be studied in more detail elsewhere.

In this paper, BAU has been evaluated in the model originally proposed in Ref.~\cite{ref: AKS}, where Majorana masses of neutrinos are generated via three-loop diagrams composed of additional scalar bosons including the DM candidate which is odd under an unbroken $Z_2$ symmetry. In order for the model to include multiple CP-violating phases, we have not imposed the softly broken $Z_2$ symmetry imposed in the original model to avoid the FCNC at tree level. Instead, for simplicity, we have assumed the flavor alignment structure in the Yukawa interactions. We have also simply assumed the alignment structure in the Higgs potential so that the Higgs couplings coincide with those in the SM at tree level. Under these phenomenological simplifications, the model still contains multiple CP-violating phases. By using destructive interferences among them, it is compatible with the stringent constraint from the EDM measurements to generate the observed baryon asymmetry along with the scenario of EWBG. We have shown a benchmark scenario which can explain neutrino mass, DM and BAU simultaneously and can satisfy all the other available experimental data. Some phenomenological predictions of the model have also been discussed.

\section*{Acknowledgement}
The works of M. A. and S. K. were supported in part by the JSPS KAKENHI Grant No.~20H00160. The work of K.~E. was supported in part by the JSPS KAKENHI Grant No.~21J11444 and the National Research Foundation of Korea (NRF) grant funded by the Korea government (MSIT) No. NRF-2021R1A2C2009718. 

\appendix

\section{Neutrino mass matrix formula}
\label{app: neutrino_formula}

In this appendix, we show the derivation of the neutrino mass formula in Eq.~(\ref{eq: formula_neutrino_mass}). 
The left diagram in Fig.~\ref{fig: Neutrino_mass_diagram} is evaluated as
\beq
i\Sigma_{ij}^1 = - i \left(\frac{ 1 }{ 16 \pi^2 } \right)^3 \kappa^2 \zeta_{\ell^i}^{\ast}  m_{\ell^i}^{} \zeta_{\ell^j}^{\ast} m_{\ell^j}^{}
\sum_{\alpha =1}^3 h^\alpha_i h^\alpha_j m_{N^\alpha}^{} F_{1\alpha}. 
\eeq
The loop function $F_{1\alpha}$ is given by 
\bal
\label{eq: F1a_start}
F_{1\alpha} = & -4i (16 \pi^2)^3 
\int \frac{ \mathrm{d}^4 p }{ (2\pi)^4} \int \frac{ \mathrm{d}^4 q }{ (2\pi)^4}
\int \frac{ \mathrm{d}^4 k }{ (2\pi)^4}\, 
\frac{ 1 }{ p^2 }
\frac{ 1 }{ q^2}
\frac{ 1 }{ p^2 - m_{H^\pm}^2 }
\frac{ 1 }{ q^2 - m_{H^\pm}^2 }
\nonumber \\[5pt] 
& \hspace{30pt} \times \frac{ 1 }{ k^2 - m_\eta^2 }
\frac{ 1 }{ k^2 - m_{N^\alpha}^2 }
\frac{ 1 }{ (k-p)^2 - m_S^2 }
\frac{ 1 }{ (k-q)^2 - m_S^2 }
\cancel{q}\cancel{p}, 
\eal
where we neglect the masses of the charged leptons in the internal lines.
Considering the interchange of momenta $p^\mu$ and $q^\mu$, 
the gamma matrices $\cancel{q}\cancel{p}$ can be replaced by $q \cdot p = q_\mu p^\mu$. 
 
The integral by $k^\mu$ can be performed by using Feynman parametrization as follows; 
\bal
& \int \frac{ \mathrm{d}^4 k }{ (2\pi)^4}
\frac{ 1 }{ k^2 - m_\eta^2 }
\frac{ 1 }{ k^2 - m_{N^\alpha}^2 }
\frac{ 1 }{ (k-p)^2 - m_S^2 }
\frac{ 1 }{ (k-q)^2 - m_S^2 } \nonumber \\[5pt]
= & \int_0^1 \tilde{\mathrm{d}}^4x \int \frac{ \mathrm{d}^4 k }{ (2\pi)^4}
\frac{ 3! }{ (k^2 - D)^4 }
= \frac{ i }{ 16 \pi^2 }  \int_0^1 \tilde{\mathrm{d}}^4x \frac{ 1 }{ D^2 }, 
\eal
where the integral $\int_0^1 \tilde{\mathrm{d}}^4x$ is defined as 
\beq 
\int_0^1 \tilde{\mathrm{d}}^4x \equiv \int^1_0 \mathrm{d}x \int^1_0 \mathrm{d}y \int^1_0 \mathrm{d}z  \int^1_0 \mathrm{d}\omega \, \delta(1-x-y-z-\omega), 
\eeq
and $D$ is given by
\beq
D = x m_{N^\alpha}^2 + (y+z) m_S^2 + \omega m_\eta^2 
- y ( 1 - y ) p^2 - z (1-z) q^2 + 2 y z p \cdot q. 
\eeq
After the Wick rotations, 
we define the Euclidean momenta $\vec{P} = (P^0, P^1, P^2, P^3)$ and 
$\vec{Q} = (Q^0, Q^1, Q^2, Q^3)$ with
\beq
P^0 = - i p^0,  \quad P^k = p^k, \quad Q^0 = - i q^0, \quad Q^k = q^k \quad (k=1,2,3). 
\eeq
Then, the loop function $F_{1\alpha}$ is given by
\bal
\label{eq: F1a_Cartesian}
F_{1\alpha} = \frac{ 4 }{ \pi^4 } & \int_0^1 \tilde{\mathrm{d}}^4x
 \int \mathrm{d}^4 P \int \mathrm{d}^4 Q\, 
\frac{ 1 }{ |\vec{P}|^2 } \frac{ 1 }{ |\vec{Q}|^2 } \frac{ 1 }{ |\vec{P}|^2 + m_{H^\pm}^2 } \frac{ 1 }{ |\vec{Q}|^2 + m_{H^\pm}^2 }
\nonumber \\
& \times \frac{ \vec{P} \cdot \vec{Q} }{ (x m_{N^\alpha}^2 + (y+z) m_S^2 + \omega m_\eta^2 + y (1-y) |\vec{P}|^2 + z (1-z ) |\vec{Q}|^2 - 2 y z \vec{P} \cdot \vec{Q})^2 }. 
\eal
The inner product $\vec{P} \cdot \vec{Q}$ can be evaluated as $\vec{P} \cdot \vec{Q} = |\vec{Q}|  P^0$ 
by arranging the $P^0$ axis along the vector $\vec{Q}$. 

The following polar coordinate is useful for the integral; 
\beq
\vec{P} = \sqrt{u}
	(\cos \theta \cos \phi, \cos \theta \sin \phi, \sin \theta \cos \chi, \sin \theta \sin \chi), 
\eeq
where variables are in the domains $0 \leq \sqrt{u} \leq \infty$, 
$0 \leq \theta \leq \pi / 2$, $0 \leq \phi \leq 2 \pi$ and $0 \leq \chi \leq 2 \pi$. 
Then, the integral by $\vec{P}$ and $\vec{Q}$ is transformed as 
\beq
\int \mathrm{d}^4 P \int \mathrm{d}^4 Q = 
\frac{ \pi^2 }{ 2 } \int_0^\infty \mathrm{d} u \int_0^\infty \mathrm{d} v 
\int_0^{2\pi} \mathrm{d}\phi \int_0^{2\pi} \mathrm{d} \chi \int_0^1 \mathrm{d} t \, 
u v t, 
\eeq
where $v \equiv |\vec{Q}|^2$, and $t \equiv \cos \theta$. 
The loop function $F_{1\alpha}$ is given by
\beq
F_{1\alpha} = \frac{ 4 }{ \pi } \int_0^1 \tilde{\mathrm{d}}^4x 
\int_0^\infty \mathrm{d} u \int_0^\infty \mathrm{d} v 
\int_0^{2\pi} \mathrm{d}\phi \int_0^1 \mathrm{d} t
\frac{ \sqrt{uv} t^2 }{ (u + m_{H^\pm}^2)( v + m_{H^\pm}^2) } 
\frac{  \cos \phi }{ ( a_{1}^{} - b_{1}^{} t \cos \phi )^2 }, 
\eeq
where 
\bal
& a_{1} = (y+z)m_S^2 + x m_{N^\alpha}^2 + \omega m_\eta^2  + z (1-z)u + y(1-y)v, \\ 
& b_{1}^{} = 2 y z \sqrt{uv}. 
\eal
In the domain of the integral, $a_{1}^{}$ is always positive and satisfies the following inequality;  
\beq
0 \leq \frac{ |b_{1}^{}| t }{ a_{1}^{} } < 1.
\eeq
Thus, the integral by $\phi$ can be performed by using the residue theorem;  
\beq
F_{1\alpha} = 8 \int_0^1 \tilde{\mathrm{d}}^4x 
\int_0^\infty \mathrm{d} u \int_0^\infty \mathrm{d} v 
\frac{ \sqrt{uv} b_1^{}  }{ (u + m_{H^\pm}^2)(v + m_{H^\pm}^2) } 
\int_0^1 \mathrm{d} t \frac{ t^3 }{ \left(a_{1}^2 - b_{1}^2 t^2\right)^{3/2} }.  
\eeq
The integral by $t$ can be easily calculated by changing the integral variable as $s= t^2$. 
Consequently, we obtain the formula in Eq.~(\ref{eq: formula_Fna}); 
\beq
\label{eq: app_F1}
 F_{1\alpha} = \int_0^1 \tilde{\mathrm{d}}^4x 
\int_0^\infty \mathrm{d} u \int_0^\infty \mathrm{d} v 
\frac{ 8\sqrt{uv} \tilde{F}(a_1^{}, b_1^{})}{ (u + m_{H^\pm}^2)(v + m_{H^\pm}^2) }, 
\eeq
where 
\beq
\tilde{F}(x,y) =   \frac{ 1 }{ y^3 }
\Bigl( \sqrt{ x^2 - y^2 } + \frac{ x^2 }{ \sqrt{x^2 - y^2 } } - 2 x \Bigr). 
\eeq

The right figure in Fig.~\ref{fig: Neutrino_mass_diagram} can be calculated in the same way. 
\beq
i\Sigma_{ij}^2 = - i \left(\frac{ 1 }{ 16 \pi^2 } \right)^3 \kappa^2 \zeta_{\ell^i}^{}  m_{\ell^i}^{} \zeta_{\ell^j}^{} m_{\ell^j}^{}
\sum_{\alpha =1}^3 h^\alpha_i h^\alpha_j m_{N^\alpha}^{} F_{2\alpha}. 
\eeq
The loop function $F_{2\alpha}$ is given by replacing $\tilde{F}(a_1^{}, b_1^{})$ in Eq.~(\ref{eq: app_F1}) into $\tilde{F}(a_2^{}, b_2^{})$ with 
\bal
& a_2^{} = (y+z) m_S^2 + x m_{N^\alpha}^2 + \omega m_\eta^2 
	+ (y+\omega)(x+z)u + (x+y)(z+\omega)v, \\
& b_2^{} = 2 (yz - x\omega) \sqrt{uv}.
\eal
Then, neutrino mass $(M_\nu)_{ij} =  i \bigl( i \Sigma_{ij}^1 + i \Sigma_{ij}^2 \bigr)$ is given by 
\beq
(M_\nu)_{ij} = \frac{ \kappa^2 (\zeta_{\ell^i}^\ast m_{\ell^i} ) (\zeta_{\ell^j}^\ast m_{\ell^j})  }{ (16 \pi^2 )^3} 
\sum_{\alpha = 1}^3 h^\alpha_i h^\alpha_j m_{N^\alpha}
\Bigl( F_{1\alpha} + F_{2\alpha} \Bigr). 
\eeq
This is the formula in Eq.~(\ref{eq: formula_neutrino_mass}).

Next, we derive the simple formula for $F_{1\alpha}$ in Eq.~(\ref{eq: F1_simple_formula}). 
First, we change the integral variables in Eq.~(\ref{eq: F1a_start}) as follows; 
\bal
F_{1\alpha}  = & -4i (16\pi^2)^3
 \int \frac{ \mathrm{d}^4 p }{ (2\pi)^4} \int \frac{ \mathrm{d}^4 q }{ (2\pi)^4}
\int \frac{ \mathrm{d}^4 k }{ (2\pi)^4}
\frac{ 1 }{ p^2 - m_{N^\alpha}^2 }
\frac{ 1 }{ p^2 - m_\eta^2 }
\nonumber \\
& \times 
\frac{ 1 }{ q^2 - m_{H^\pm}^2 }
\frac{ 1 }{ q^2 }
\frac{ 1 }{ (p+q)^2 - m_S^2 }
\frac{ 1 }{ k^2 - m_{H^\pm}^2 }
\frac{ 1 }{ k^2 }
\frac{ 1 }{ (p+k)^2 - m_S^2 }
(q \cdot k), 
\eal
where the gamma matrices $\cancel{q}\cancel{k}$ is replaced into $q \cdot k$ since the integrand is symmetric with respect to $q \leftrightarrow k$. 
Then, $F_{1\alpha}$ can be calculated as follows; 
\bal
F_{1\alpha} = & \frac{ 4 (16\pi^2)^3 }{ i m_{H^\pm}^4 } 
\int \frac{ \mathrm{d}^4 p }{ (2\pi)^4} 
\frac{ 1 }{ p^2 - m_{N^\alpha}^2 }
\frac{ 1 }{ p^2 - m_\eta^2 }
\int \frac{ \mathrm{d}^4 q }{ (2\pi)^4}
\left( \frac{ 1 }{ q^2 - m_{H^\pm}^2 }- \frac{ 1 }{ q^2 } \right)
\frac{ q^\mu }{ (p+q)^2 - m_S^2 }
\nonumber \\
& \times 
\int \frac{ \mathrm{d}^4 k }{ (2\pi)^4}
\left( \frac{ 1 }{ k^2 - m_{H^\pm}^2 } - \frac{ 1 }{ k^2 } \right)
\frac{ k_\mu }{ (p+k)^2 - m_S^2 }
\nonumber \\[10pt]
= & - \frac{ 64 \pi^2 }{ i m_{H^\pm}^4 }
\int \frac{ \mathrm{d}^4 p }{ (2\pi)^4} 
\frac{ p^2 }{ p^2 - m_{N^\alpha}^2 }
\frac{ 1 }{ p^2 - m_\eta^2 } 
\Bigl( B_1(m_{H^\pm}^2, m_S^2; p^2) - B_1(0, m_S^2; p^2) \Bigr)^2, 
\eal
where $B_1$ is the Passarino-Veltman function for one-loop integrals~\cite{Passarino:1978jh}. 
By using the Wick rotation, we obtain the following result; 
\bal
F_{1\alpha} = & \frac{ 4 }{ m_{H^\pm}^4 (m_{N^\alpha}^2 - m_\eta^2)} 
\int_0^\infty \mathrm{d}x\, x \left( \frac{ m_{N^\alpha}^2 }{ x + m_{N^\alpha}^2 } - \frac{ m_\eta^2 }{ x + m_\eta^2 } \right)
\nonumber \\
& \times \Bigl( B_1(m_{H^+}^2, m_S^2; -x) - B_1(0, m_S^2; -x) \Bigr)^2. 
\eal
This formula is the same with that in Eq.~(\ref{eq: F1_simple_formula}).

\section{The tensor structure of the scattering amplitude for $\eta \eta \to \gamma \gamma$}
\label{app: 2Eta2gamma}

In this appendix, we show the cross section for the diphoton annihilation of the dark matter $\eta \eta \to \gamma \gamma$. 
Before calculating Feynman diagrams in Fig.~\ref{fig: diphoton}, 
we give a general discussion on the scattering amplitude for the diphoton annihilation of a pair of real scalar bosons.  
In the following, the momenta of scalar bosons are denoted by $p_1^\mu$ and $p_2^\mu$, while those of photons are denoted by $k_1^\mu$ and $k_2^\mu$. 

The scattering amplitude of $\eta \eta \to \gamma \gamma$ are generally given by 
\begin{equation}
i \mathcal{M} = \frac{ i \alpha }{ 4 \pi } \mathcal{M}^{\mu\nu}
	\epsilon^\ast_\mu(k_1) \epsilon^\ast_\nu (k_2), 
\end{equation}
where $\epsilon_\mu(k)$ is a polarization vector of the photon with momentum $k^\mu$, and $\alpha$ is the fine structure constant. The prefactor $\alpha/(4\pi)$ is just for normalization. 
Without parity-violating interactions, 
the tensor structure of $\mathcal{M}^{\mu\nu}$ is given by 
\begin{align}
\mathcal{M}^{\mu\nu} = 
	& \ a_{00}^{} g^{\mu\nu} + b_{12}^{} k_1^\nu k_2^\mu 
	+ c_{11} k_1^\nu p_1^\mu 
	+ c_{12} k_1^\nu p_2^\mu + c_{21} k_2^\mu p_1^\nu 
	+ c_{22} k_2^\mu p_2^\nu 
	\nonumber \\
	& + d_{11} p_1^\mu p_1^\nu 
	+ d_{12} p_1^\mu p_2^\nu + d_{21} p_2^\mu p_1^\nu 
	+ d_{22} p_2^\mu p_2^\nu, 
\end{align}
where we use $\epsilon_\mu^\ast (k_1) k_1^\mu = \epsilon_\nu^\ast(k_2) k_2^\nu = 0$. 
The coefficients $a_{00}^{}$, $b_{12}^{}$, $c_{ij}^{}$ and $d_{ij}^{}$ ($i,j=1,2$) are functions of the Mandelstam variables. 
The scattering amplitude $\mathcal{M}^{\mu\nu}$ has to be symmetric for the replacements $p_1^{} \leftrightarrow p_2^{}$ and $(k_1^{}, \mu) \leftrightarrow (k_2, \nu)$. 
This leads to the following conditions 
\begin{equation}
a_{00}^{} = \tilde{a}_{00}^{}, \quad 
b_{12}^{} = \tilde{b}_{12}^{}, \quad 
c_{11}^{} = c_{22} = \tilde{c}_{12}^{} = \tilde{c}_{21}^{}, \quad 
d_{11}^{} = d_{22}^{}, \quad
d_{12} = \tilde{d}_{21}, 
\end{equation}
where functions with a tilde represent the functions with the replacement of Mandelstam variables $t \leftrightarrow u$. 
In addition, by the momentum conservation, the terms $p_1^\mu p_2^\nu$ and $p_2^\mu p_1^\nu$ can be represented by the other tensors. 
Consequently, $\mathcal{M}^{\mu\nu}$ are generally given by 
\begin{equation}
\label{eq: CF_abcd}
\mathcal{M}^{\mu\nu} = a g^{\mu\nu} + b k_1^\nu k_2^\mu
	+ c \bigl( k_1^\nu p_1^\mu + k_2^\mu p_2^\nu \bigr)
	+ \tilde{c} \bigl( k_1^\nu p_2^\mu + k_2^\mu p_1^\nu \bigr)
	+ d \bigl( p_1^\mu p_1^\nu + p_2^\mu p_2^\nu \bigr). 
\end{equation}
The coefficients $a$, $b$ and $d$ are symmetric for $t \leftrightarrow u$. 
On the other hand, $c$ are generally not symmetric. 

Four coefficient functions can be reduced to two by using the Ward--Takahashi (WT) identity, $\mathcal{M}^{\mu\nu} k_{1 \mu} = \mathcal{M}^{\mu\nu} k_{2 \nu} = 0$. 
The WT identity gives the two conditions for the coefficient functions;
\begin{equation}
\label{eq: WT_identity}
\left\{
\begin{array}{l}
a + b(k_1 \cdot k_2) + c (k_1\cdot p_1) + \tilde{c} (k_1 \cdot p_2) = 0, \\
c (k_1 \cdot k_2) + d (k_1 \cdot p_2) = 0. 
\end{array}
\right.
\end{equation}
Thus, the functions $a$ and $c$ can be represented by using $b$ and $d$. 
Finally, the scattering amplitude $\mathcal{M}^{\mu\nu}$ can be represented by only two functions; 
\begin{align}
\mathcal{M}^{\mu\nu} = & \ A_{\gamma \gamma}^{}
	\bigl( s g^{\mu\nu} - 2 k_1^\nu k_2^\mu \bigr)
	\nonumber \\
	& + B_{\gamma \gamma}^{} 
	\Bigl\{
		(u-m_\eta^2)(t-m_\eta^2) g^{\mu\nu}
		+ s \bigl(p_1^\mu p_1^\nu + p_2^\mu p_2^\nu \bigr)
		\nonumber \\
		& \hspace{40pt} 
		+ (u-m_\eta^2) \bigl( k_1^\nu p_1^\mu + k_2^\mu p_2^\nu \bigr)
		+ (t - m_\eta^2) \bigl( k_1^\nu p_2^\mu + k_2^\mu p_1^\nu \bigr)
	\Bigr\}, 
\end{align}
where $A_{\gamma \gamma} = - b/2$ and $B_{\gamma \gamma} = d / s$. 
The scattering cross section is then given by
\begin{align}
\sigma_{\eta \eta \to \gamma \gamma} 
	= \frac{ \alpha^2 }{ 512 \pi^3 } 
	\frac{ 1 }{ \sqrt{ s ( s - 4m_\eta^2) } }
	\int_{-1}^1 \mathrm{d}\cos \theta\, 
	\Bigl(
		& s^2 |A_{\gamma\gamma}|^2 
		+ s^2 m_\eta^2 
			\bigl( A_{\gamma \gamma} B^\ast_{\gamma \gamma}
				+ A_{\gamma \gamma}^\ast B_{\gamma \gamma} \bigr)
			\nonumber \\
			& + |B_{\gamma \gamma} |^2 
			\bigl( s^2 m_\eta^4 + (ut - m_\eta^4)^2 \bigr)
	\Bigr), 
\end{align}
which leads Eq.~(\ref{eq: diphoton_annihi}). 
This result is so general that we can use it for a diphoton annihilation for any real scalar boson $\eta$ which does not have to be a dark matter as long as the parity-violating interactions can be neglected. 

Next, we show the calculation of $\eta \eta \to \gamma \gamma$ in the model described by Feynman diagrams in Fig.~\ref{fig: diphoton}. 
To this end, the formulae for $A_{\gamma \gamma}$ and $B_{\gamma \gamma}$ are required. 
They are given by the sum of two kinds of contributions; 
\begin{equation}
\label{eq: Diphoton_function_in_AKS}
A_{\gamma \gamma} = A_{\gamma\gamma}^1 + A_{\gamma \gamma}^2, \quad
B_{\gamma \gamma} = B_{\gamma\gamma}^1 + B_{\gamma \gamma}^2, 
\end{equation} 
where $A_{\gamma\gamma}^1$ and $B_{\gamma \gamma}^1$ are proportional to the scalar coupling $\kappa$, 
and $A_{\gamma\gamma}^2$ and $B_{\gamma \gamma}^2$ are proportional to other scalar couplings $\sigma_2$ and $\xi$. 

First, we consider the Feynman diagrams induced by $\kappa$, which are the upper-left, upper-middle and right diagrams in Fig.~\ref{fig: diphoton}. 
$A_{\gamma \gamma}^1$ and $B_{\gamma \gamma}^1$ are given by 
\begin{align}
& A_{\gamma \gamma}^1 
	= 8 (\kappa v)^2 
	\Bigl(  D_{13}(\bm{d1}) + D_{13}(\bm{\tilde{d1}})
		+ D_{23}(\bm{d1}) + D_{23}(\bm{\tilde{d1}}) 
		+ D_{13}(\bm{d2}) + D_{13}(\bm{\tilde{d2}})
		\nonumber \\
		& \hspace{75pt}
		+ D_{23}(\bm{d2}) + D_{23}(\bm{\tilde{d2}})
		+ D_{23}(\bm{d3}) + D_{23}(\bm{\tilde{d3}})
	\Bigr), \\
& B_{\gamma \gamma}^1 
	= \frac{ 8 }{ s } (\kappa v)^2 
	\Bigl( D_{22}(\bm{d1}) + D_{22}(\bm{\tilde{d1}}) 
		+ D_{22}(\bm{d2}) + D_{22}(\bm{\tilde{d2}})
		\nonumber \\
		& \hspace{75pt} 
		+ D_{11}(\bm{d3}) + D_{11}(\bm{\tilde{d3}}) 
		+ 2 D_{12}(\bm{d3}) + 2 D_{12}(\bm{\tilde{d3}}) 
		\nonumber \\
		& \hspace{75pt} 
		+ D_{22}(\bm{d3}) + D_{22}(\bm{\tilde{d3}}) 
		+ D_{1}(\bm{d3}) + D_{1}(\bm{\tilde{d3}}) 
		+ D_{2}(\bm{d3}) + D_{2}(\bm{\tilde{d3}}) 
	\Bigr), 
\end{align} 
where $D_{ij}$ ($i,j=1,2,3$ and $i \leq j$) and $D_i$ ($i=1,2$) are the Passarino-Veltman functions for one-loop integrals with four external lines~\cite{Passarino:1978jh}. 
The arguments of the functions are abbreviated by $\bm{d1}$, $\bm{d2}$ and $\bm{d3}$ as follows; 
\begin{align}
& \bm{d1} = (0, m_\eta^2, m_\eta^2, 0, t, s; m_{H^\pm}^2, m_{H^\pm}^2, m_S^2, m_{H^\pm}^2), \\
& \bm{d2} = (0, m_\eta^2, m_\eta^2, 0, t, s; m_S^2, m_S^2, m_{H^\pm}^2, m_S^2), \\
& \bm{d3} = (m_\eta^2, 0, m_\eta^2, 0, t, u; m_{H^\pm}^2, m_S^2, m_S^2, m_{H^\pm}^2),  
\end{align}
where the notation follows \texttt{LoopTools}~\cite{Hahn:1998yk}. 
The arguments with tilde $\tilde{\bm{d1}}$, $\tilde{\bm{d2}}$ and $\tilde{\bm{d3}}$ represents the replacement $t \leftrightarrow u$ in $\bm{d1}$, $\bm{d2}$ and $\bm{d3}$, respectively.

Second, we consider the lower-left and lower-middle diagrams in Fig.~\ref{fig: diphoton}. 
The diagrams including a loop of $H^\pm$ and $S^\pm$ are proportional to $\sigma_2$ and $\xi$, respectively. 
$A_{\gamma \gamma}^2$ and $B_{\gamma \gamma}^2$ in Eq.~(\ref{eq: Diphoton_function_in_AKS}) are given by 
\begin{align}
& A_{\gamma \gamma}^2 = 
	2 \sigma_2 
	\left\{ 1 - G\left( \frac{ s }{ 4 m_{H^\pm}^2 } \right) \right\}
	+ 2 \xi \left\{ 1 - G\left( \frac{ s }{ 4 m_S^2 } \right)
	\right\}, \\
& B_{\gamma \gamma}^2 = 0, 
\end{align}
where the function $G(r)$ is defined as 
\begin{equation}
G(r) = \left\{
\begin{array}{ll}
\displaystyle{- \frac{ 1 }{ 4 r }
	\left\{ 
		\log \left( \frac{ 1 + \sqrt{1 - r^{-1}} }{ 1 - \sqrt{ 1 - r^{-1} } } \right)
		- i \pi 
	\right\}^2}\quad  & (r > 1), \\[17pt]
\displaystyle{\frac{ 1 }{ r } \Bigl( \mathrm{sin}^{-1} \sqrt{r} \Bigr)^2 }
	& (0 < r < 1). 
\end{array}
\right.
\end{equation}
$B_{\gamma \gamma}^2 = 0$ is the consequence of the fact that the Feynman diagrams depend on only $(p_1 +p_2)^\mu,$ $k_1^\mu$ and $k_2^\mu$ but not on $p_1^\mu$ and $p_2^\mu$ separately. 
Then, the amplitude can be represented by only $A_{\gamma \gamma}$ like in the case of the Higgs diphoton decay~\cite{Ellis:1975ap, Shifman:1979eb}. 

We now obtain the all required formulae to evaluate $\eta \eta \to \gamma \gamma$ in the model. 
Although, we here show only $A_{\gamma \gamma}$ and $B_{\gamma \gamma}$, we have calculated all the coefficient functions $a$, $b$, $c$ and $d$ in Eq.~(\ref{eq: CF_abcd}), 
and we have confirmed the WT identity in Eq.~(\ref{eq: WT_identity}) numerically with at least $10^{-10}$ accuracy.

\section{Formulae for field-dependent masses} 
\label{app: field-dependent_mass}

In this appendix, 
the formulae for the field-dependent masses used in Sec.~\ref{sec: EWPT} are shown. 
We show formulae for $\hat{m}^2_a$ in the following. 
By taking $T=0$, the field-dependent masses at zero-temperature $\tilde{m}^2_a$ are obtained. 
The classical field for the Higgs doublets is defined as follows; 
\beq
\Phi_1^\mathrm{cl} = \frac{ 1 }{ \sqrt{2} }
\begin{pmatrix}
0 \\
\varphi_1 \\
\end{pmatrix}, \quad 
\Phi_2^\mathrm{cl} = \frac{ 1 }{ \sqrt{2} } 
\begin{pmatrix}
0 \\
\varphi_2 + i \varphi_3 \\
\end{pmatrix}. 
\eeq
In the following results, we assume the alignment in the Higgs potential; $\lambda_6 = \mu_3^2 = 0$. 
In addition, we consider the case that $\lambda_5$ is a real number by an appropriate rephasing of $\Phi_2$. 

First, the field-dependent masses for $G^\pm$ and $H^\pm$ are given by the eigenvalues of a $2\times 2$ hermitian matrix $\hat{M}_\pm^2$. 
Each element of $\hat{M}_\pm^2$ is given by 
\bal
& (\hat{M}_\pm^2)_{\scalebox{0.6}{$G^+G^-$}} = \mu_1^2 + \frac{ \lambda_1 }{ 2 } \varphi_1^2 
	+ \frac{ \lambda_3 }{ 2 } ( \varphi_2^2 + \varphi_3^2 ) \nonumber \\
	& \hspace{60pt}
	+ \frac{ T^2 }{ 24 } 
	\Bigl( 
		6 \lambda_1 + 4 \lambda_3 + 2 \lambda_4 + 2 \rho_1 + \sigma_1 
		+ 6 y_t^2 + \frac{ 9 g_L^2 }{ 2 } + \frac{ 3 g_Y^2 }{ 2 }
	\Bigr), \\[10pt]
& (\hat{M}_\pm^2)_{\scalebox{0.6}{$G^+H^-$}}  
	= (\hat{M}_\pm^2)_{\scalebox{0.6}{$H^+G^-$}}^\ast
	= \frac{ 1 }{ 2 } (\lambda_4 + \lambda_5 ) \varphi_2 
	+ \frac{ i }{ 2 } ( \lambda_4 - \lambda_5 ) \varphi_3 
	+ \frac{ \lambda_7^\ast }{ 2 } (\varphi_2^2 + \varphi_3^2) \nonumber \\
	& \hspace{120pt}
	+ \frac{ T^2 }{ 24 } 
	\Bigl( 
		6 \lambda_7^\ast + 2 \rho_{12} + \sigma_{12} + 6 y_t^2 \zeta_u^\ast
	\Bigr), \\[10pt]
& (\hat{M}_\pm^2)_{\scalebox{0.6}{$H^+H^-$}} = \mu_2^2 
	+  \frac{ \lambda_2 }{ 2 } (\varphi_2^2 + \varphi_3^2 )
	+ \frac{ \lambda_3 }{ 2 } \varphi_1^2 
	+ \mathrm{Re}[\lambda_7] \varphi_1 \varphi_2 
	- \mathrm{Im}[\lambda_7] \varphi_1 \varphi_3 \nonumber \\
	& \hspace{60pt}
	+ \frac{ T^2 }{ 24 } 
	\Bigl( 
		6 \lambda_2 + 4\lambda_3 + 2 \lambda_4 + 2 \rho_2 + \sigma_2 + 6 y_t^2 |\zeta_u|^2
		+ \frac{ 9 g_L^2 }{ 2 } + \frac{ 3 g_Y^2 }{ 2 }
	\Bigr), 
\eal
where $y_t$ is defined as $y_t = \sqrt{2} m_t / v$. 
Next, the field-dependent masses for the neutral scalar bosons $H_1$, $H_2$, $H_3$, and $G^0$ are given by the eigenvalues of a $4\times 4$ real symmetric matrix $\hat{M}_0^2$. 
Each element of $\tilde{M}_0^2$ is given by
\bal
& (\hat{M}_0^2)_{\scalebox{0.6}{$G^0G^0$}}
	= \mu_1^2 + \frac{ \lambda_1 }{ 2 } \varphi_1^2 
	+ \frac{ 1 }{ 2 }( \lambda_3 + \lambda_4 - \lambda_5 ) \varphi_2^2 
	+ \frac{ 1 }{ 2 } ( \lambda_3 + \lambda_4 + \lambda_5 ) \varphi_3^2
	\nonumber \\
	& \hspace{60pt} 
	+ \frac{ T^2 }{ 24 } 
	\Bigl(
		6 \lambda_1 + 4 \lambda_3 + 2 \lambda_4 + 2 \rho_1 + \sigma_1 
		+ 6 y_t^2 + \frac{ 9 g_L^2 }{ 2 } + \frac{ 3 g_Y^2 }{ 2 } 
	\Bigr), \\[10pt]
& (\hat{M}_0^2)_{\scalebox{0.6}{$H_1H_1$}}
	= \mu_1^2 + \frac{ 3 }{ 2 } \lambda_1 \varphi_1^2 
	+ \frac{ 1 }{ 2 }(\lambda_3 + \lambda_4 + \lambda_5 ) \varphi_2^2 
	+ \frac{ 1 }{ 2 }( \lambda_3 + \lambda_4 - \lambda_5 ) \varphi_3^2
	\nonumber \\
	& \hspace{60pt}
	+ \frac{ T^2 }{ 24 } 
	\Bigl(
		6 \lambda_1 + 4 \lambda_3 + 2 \lambda_4 + 2 \rho_1 + \sigma_1
		+ 6 y_t^2 + \frac{ 9 g_L^2 }{ 2 } + \frac{ 3 g_Y^2 }{ 2 } 
	\Bigr), 
\eal

\bal
& (\hat{M}_0^2)_{\scalebox{0.6}{$H_2H_2$}}
	= \mu_2^2 + \frac{ \lambda_2 }{ 2 } ( 3 \varphi_2^2 + \varphi_3^2 )
	+ \frac{ 1 }{ 2 }(\lambda_3 + \lambda_4 + \lambda_5 )\varphi_1^2 
	+ 3 \mathrm{Re}[\lambda_7] \varphi_1 \varphi_2 
	- \mathrm{Im}[\lambda_7] \varphi_1 \varphi_3 \nonumber \\
	& \hspace{60pt}
	+ \frac{ T^2 }{ 24 }
	\Bigl(
		6 \lambda_2 + 4 \lambda_3 + 2 \lambda_4 + 2 \rho_2 + \sigma_2
		+ 6 y_t^2 |\zeta_u|^2 + \frac{ 9 g_L^2 }{ 2 } + \frac{ 3 g_Y^2 }{ 2 }
	\Bigr), \\[10pt]
& (\hat{M}_0^2)_{\scalebox{0.6}{$H_3H_3$}}
	= \mu_2^2 + \frac{ \lambda_2 }{ 2} ( 3 \varphi_2^2 + \varphi_3^2 )
	+ \frac{ 1 }{ 2 }(\lambda_3 + \lambda_4 -\lambda_5 ) \varphi_1^2
	+ \mathrm{Re}[\lambda_7]\varphi_1\varphi_2
	- 3 \mathrm{Im}[\lambda_7] \varphi_1 \varphi_3 \nonumber \\
	& \hspace{60pt}
	+ \frac{ T^2 }{ 24 }
	\Bigl(
		6 \lambda_2 + 4 \lambda_3 + 2 \lambda_4 + 2 \rho_2 + \sigma_2
		+ 6 y_t^2 |\zeta_u|^2 + \frac{ 9 g_L^2 }{ 2 } + \frac{ 3 g_Y^2 }{ 2 }
	\Bigr), \\[10pt]
& (\hat{M}_0^2)_{\scalebox{0.6}{$G^0 H_1$}} 
	= (\hat{M}_0^2)_{\scalebox{0.6}{$H_1G^0$}}
	= \lambda_5 \varphi_2 \varphi_3, \\[10pt]
& (\hat{M}_0^2)_{\scalebox{0.6}{$G^0H_2$}}
	= (\hat{M}_0^2)_{\scalebox{0.6}{$H_2G^0$}}
	= \lambda_5 \varphi_1 \varphi_3 + \mathrm{Re}[\lambda_7] \varphi_2 \varphi_3 
	+ \frac{ 1 }{ 2 } \mathrm{Im}[\lambda_7] (3 \varphi_2^2 +\varphi_3^2 )
	\nonumber \\
	& \hspace{120pt}
	+ \frac{ T^2 }{ 24 }
	\Bigl( 
		6 \mathrm{Im}[\lambda_7] + 2 \mathrm{Im}[\rho_{12}]
		+ \mathrm{Im}[\sigma_{12}] - 6 y_t^2 \mathrm{Im}[\zeta_u]
	\Bigr), \\[10pt]
& (\hat{M}_0^2)_{\scalebox{0.6}{$G^0H_3$}}
	= (\hat{M}_0^2)_{\scalebox{0.6}{$H_3G^0$}}
	= \lambda_5 \varphi_1 \varphi_2 
	+ \frac{ 1 }{ 2 } \mathrm{Re}[\lambda_7](\varphi_2^2 + 3 \varphi_3^2)
	+ \mathrm{Im}[\lambda_7]\varphi_2 \varphi_3
	\nonumber \\
	& \hspace{120pt}
	+ \frac{ T^2 }{ 24 }
	\Bigl(
		6 \mathrm{Re}[\lambda_7] + 2 \mathrm{Re}[\rho_{12}] + \mathrm{Re}[\sigma_{12}] + 6 y_t^2 \mathrm{Re}[\zeta_u]
	\Bigr), \\[10pt]
& (\hat{M}_0^2)_{\scalebox{0.6}{$H_1H_2$}}
	= (\hat{M}_0^2)_{\scalebox{0.6}{$H_2H_1$}}
	= (\lambda_3 + \lambda_4 + \lambda_5 ) \varphi_1 \varphi_2 
	+ \frac{ 1 }{2 }\mathrm{Re}[\lambda_7] (3 \varphi_2^2 + \varphi_3^2)
	- \mathrm{Im}[\lambda_7]\varphi_2 \varphi_3 \nonumber \\
	& \hspace{120pt}
	+ \frac{ T^2 }{ 24 }
	\Bigl(
		6 \mathrm{Re}[\lambda_7] + 2 \mathrm{Re}[\rho_{12}]
		+ \mathrm{Re}[\sigma_{12}] + 6 y_t^2 \mathrm{Re}[\zeta_u]
	\Bigr), \\[10pt]
& (\hat{M}_0^2)_{\scalebox{0.6}{$H_1H_3$}} 
	= (\hat{M}_0^2)_{\scalebox{0.6}{$H_3H_1$}}
	= (\lambda_3 + \lambda_4 - \lambda_5) \varphi_1 \varphi_3 
	- \frac{ 1 }{ 2 } \mathrm{Im}[\lambda_7] ( \varphi_2^2 + 3 \varphi_3^2 )
	+ \mathrm{Re}[\lambda_7]\varphi_2 \varphi_3 \nonumber \\
	& \hspace{120pt}
	+ \frac{ T^2 }{ 24 }
	\Bigl(
		-6 \mathrm{Im}[\lambda_7] -2  \mathrm{Im}[\rho_{12}]
		- \mathrm{Im}[\sigma_{12}] + 6 y_t^2 \mathrm{Im}[\zeta_u]
	\Bigr), \\[10pt]
& (\hat{M}_0^2)_{\scalebox{0.6}{$H_2H_3$}} 
	= (\hat{M}_0^2)_{\scalebox{0.6}{$H_3H_2$}} 
	= \lambda_2 \varphi_2 \varphi_3 
	+ \mathrm{Re}[\lambda_7] \varphi_1 \varphi_3 
	- \mathrm{Im}[\lambda_7] \varphi_1 \varphi_2.
\eal
The field-dependent masses for the $Z_2$-odd scalar bosons $S^\pm$ and $\eta$ are given by 
\bal
& \hat{m}_S^2 = 
\mu_S^2 + \frac{ \rho_1 }{ 2 } \varphi_1^2 
+ \frac{ \rho_2 }{ 2 } (\varphi_2^2 + \varphi_3^2 )
+ \mathrm{Re}[\rho_{12}] \varphi_1 \varphi_2
- \mathrm{Im}[\rho_{12}]\varphi_1 \varphi_3
+ \frac{ T^2 }{ 6 } ( \rho_1 + \rho_2 + 6 g_Y^2 ), \\[10pt]
& \hat{m}_\eta^2 = 
\mu_\eta^2 + \frac{ \sigma_1 }{ 2 } \varphi_1^2 
+ \frac{ \sigma_2 }{ 2 } (\varphi_2^2 + \varphi_3^2 )
+ \mathrm{Re}[\sigma_{12}] \varphi_1 \varphi_2 
- \mathrm{Im}[\sigma_{12}]\varphi_1 \varphi_3
+ \frac{ T^2 }{ 6 } ( \sigma_1 + \sigma_2 ). 
\eal
The field-dependent masses for the gauge bosons $W_\mu^a$ ($a=1,2,3$) and $B_\mu$ are given by a $2\times 2$ real symmetric matrix $(\hat{M}^2_V)$ whose elements are defined as follows;
\bal
& (\hat{M}_V^2)_{\scalebox{0.6}{$W^a_\mu W^b_\nu$}}
= \Bigl( \frac{ g_L^2 }{4} \varphi^2 + 2 g_L^2 T^2 \delta_{\mu, ||} \Bigr) \delta_{ab} \delta_{\mu\nu}, \\[10pt]
& (\hat{M}_V^2 )_{\scalebox{0.6}{$W^3_\mu B_\nu$}}
= \frac{ 1 }{ 4 } g_L g_Y \varphi^2 \delta_{\mu\nu}, \\[10pt]
& (\hat{M}_V^2 )_{\scalebox{0.6}{$B_\mu B_\nu$}}
= \Bigl( \frac{ g_Y^2 }{4} \varphi^2 + \frac{8}{3} g_Y^2 T^2 \delta_{\mu, ||} \Bigr)\delta_{\mu\nu}, 
\eal
where $\varphi^2$ is the total VEV; $\varphi^2 = \varphi_1^2 + \varphi_2^2 + \varphi_3^2$. 
The symbol $\delta_{\mu, ||}$ means that only the longitudinal components of the gauge bosons receive the thermal corrections at this order~\cite{Espinosa:1992kf}. 

Finally, we show the field-dependent masses of the top quark. 
We do not consider the thermal resummation for the top quark since thermal fermion loops do not have the infrared problems caused by zero modes of Matsubara frequency~\cite{Dolan:1973qd}. 
Then, the field-dependent mass of the top quark is given by
\bal
\hat{m}_t^2 = \tilde{m}_t^2 = 
\frac{m_t^2}{v^2}
\Bigl\{
	\bigl(\varphi_1 + \varphi_2 |\zeta_u| \cos \theta_u 
		+ \varphi_3 | \zeta_u | \sin \theta_u \bigr)^2 &
\nonumber \\
	+ \bigl(\varphi_2 | \zeta_u |\sin \theta_u - \varphi_3 | \zeta_u | \cos \theta_u \bigr)^2 &
\Bigr\}. 
\eal

\end{document}